\documentclass[12pt]{article}
\usepackage{amsmath, amssymb}

\newtheorem{theorem}{Theorem}[section]

\newtheorem{lemma}[theorem]{Lemma}
\newtheorem{proposition}[theorem]{Proposition}
\newtheorem{definition}{Definition}[section]
\numberwithin{equation}{section}

\newcommand{\R}{\mathbb{R}}
\newcommand{\Complex}{\mathbb{C}}
\renewcommand{\Re}{\mathop{\mathrm{Re}}}
\renewcommand{\Im}{\mathop{\mathrm{Im}}}

\newcommand{\myspan}{\mathop{\rm span}}
\newcommand{\dist}{\mathop{\rm dist}}

\newcommand{\longto}{\longrightarrow}
\newcommand{\norm}[1]{\left\Vert #1 \right\Vert}
\newcommand{\length}[1]{\left| #1 \right|}
\newcommand{\bkA}[1]{\left \langle #1 \right \rangle}

\newcommand{\bke}[1]{\left( #1 \right)}
\newcommand{\bkt}[1]{\left[ #1 \right]}
\newcommand{\bket}[1]{\left\{ #1 \right\}}

\newcommand{\e}{\varepsilon}
\newcommand{\al}{\alpha}
\newcommand{\ka}{\kappa}
\newcommand{\om}{\omega}
\newcommand{\la}{\lambda}

\newcommand{\loc}{_{\mathrm{loc}}}
\newcommand{\myremark}{{\bf Remark}\quad}
\newcommand{\myproof}{\noindent {\bf Proof.}\quad}
\newcommand{\myendproof}{\hspace*{\fill}{{\bf \small Q.E.D.}} \vspace{10pt}}

\renewcommand{\L}{\mathcal{L}}

\newcommand{\Pc}{\, \mathbf{P}\! _\mathrm{c} \, \! }

\newcommand{\PcL}{\, \mathbf{P}\! _\mathrm{c} \! ^{\L_1}}

\newcommand{\Hc}{\, \mathbf{H} _\mathrm{c} \, \! }

\newcommand{\eigen}{\mathbf{E}_1}

\newcommand{\xii}{\xi_\infty}

\newcommand{\pd}{\partial}
\newcommand{\donothing}[1]{}
\newcommand{\wbar}[1]{\overline{\rule{0pt}{2.4mm} {#1}}}
\newcommand{\wt}{\widetilde}

\newcommand{\vect}[1]{\begin{bmatrix} #1 \end{bmatrix}}
\newcommand{\svect}[1]
{\bkt{\begin{smallmatrix} #1\end{smallmatrix}}}

\newcommand{\bQ}{\mathbf{Q}}

\newcommand{\Tmap}{{\bf \Omega}}
\newcommand{\RE}{\mathop{\,\mathbf{RE}\,}}

\newcommand{\evalue}{\omega_*} 
\newcommand{\evector}{\mathbf{\Phi}}
\newcommand{\gs}{\wt \phi_0}
\newcommand{\es}{\widehat Q_1}
\newcommand{\psias}{\psi_{\mathrm{as}}}

\newcommand{\PM}{ {\bf P}\! _{M_1} }

\begin{document}

\baselineskip 20pt

\vspace*{1in}

\begin{center}

{\bf \large
STABLE DIRECTIONS FOR\\ EXCITED STATES OF\\ NONLINEAR SCHR\"ODINGER
EQUATIONS}

\bigskip

{\bf Tai-Peng Tsai \ and \ Horng-Tzer Yau}

\end{center}

\bigskip

\begin{abstract}

We consider  nonlinear Schr\"odinger equations in $\R^3$. Assume
that the linear Hamiltonians have two bound states. For certain
finite codimension subset in the space of  initial data, we
construct solutions converging to the excited states in both
non-resonant and resonant cases. In the resonant case, the
linearized  operators around the excited states are non-self
adjoint perturbations to some linear Hamiltonians with embedded
eigenvalues. Although self-adjoint perturbation turns embedded
eigenvalues into resonances, this class of non-self adjoint
perturbations turn an embedded eigenvalue into two eigenvalues
with the distance to the continuous  spectrum given  to the
leading order by the  Fermi golden rule.

\end{abstract}

\section{Introduction}

Consider the nonlinear  Schr\"odinger equation
\begin{equation} \label{Sch}
i \pd _t \psi = (-\Delta + V) \psi + \la |\psi|^2 \psi, \qquad
\psi(t=0)= \psi_0
\end{equation}
where $V$ is a smooth localized potential,  $\la $ is an order one
parameter and $\psi=\psi(t,x):\R\times \R^3 \longto \Complex$ is a
wave function. The goal of this paper is to study the asymptotic
dynamics of the solution for  initial data $\psi_0$ near some {\it
nonlinear excited state}.

For any solution $\psi(t)\in H^1(\R^3)$ the $L^2$-norm
and the Hamiltonian
\begin{equation} \label{1-2}
{\cal H}[\psi] = \int \frac 12 |\nabla \psi|^2 +  \frac 12 V |\psi|^2
+ \frac 14 \la |\psi|^4 \, d x ~,
\end{equation}
are constants for all $t$. The global well-posedness for small
solutions in $H^1(\R^3)$ can be proved using these conserved
quantities and a continuity argument.

We assume that the linear Hamiltonian $H_0 :=- \Delta + V$ has two
simple eigenvalues $e_0<e_1<0$ with normalized eigen-functions
$\phi_0$, $\phi_1$.
The nonlinear bound states to the Schr\"odinger equation
\eqref{Sch} are solutions to the equation
\begin{equation}   \label{Q.eq}
    (-\Delta + V) Q + \la |Q|^2 Q = EQ  ~.
\end{equation}
They are critical points to the Hamiltonian ${\cal H}[\psi] $
defined in \eqref{1-2} subject to the constraint that the
$L^2$-norm of $\psi$ is fixed.

We may obtain two families of such bound states by standard
bifurcation theory, corresponding to the two eigenvalues of  the
linear Hamiltonian. For any $E$ sufficiently close to $e_0$ so
that $E-e_0$ and $\la$ have the same sign, there is a unique
positive solution $Q=Q_E$ to \eqref{Q.eq} which decays
exponentially  as $x \to \infty$. See Lemma 2.1 of \cite{TY2}. We
call this family the {\it nonlinear ground states} and we refer to
it as $\bket{Q_{E}}_{E}$. Similarly, there is a {\it nonlinear
excited state} family $\bket{Q_{1,E_1}}_{E_1}$. We will abbreviate them
as $Q$ and $Q_{1}$. From the same Lemma 2.1 of \cite{TY2},
these solutions are small and we
have $\norm{Q_{E}} \sim |E-e_0|^{1/2}$ and $\norm{Q_{1,E_1}} \sim
|E_1-e_1|^{1/2}$.

It is well-known that the
family of nonlinear ground states is stable in the sense that if
\[
   \inf_{\Theta, E} \norm{ \psi(t) -Q_E \, e^{i \Theta} }_{L^2}
\]
is small for $t=0$, it remains so for all $t$, see \cite{RW}. Let
$\norm{ \cdot }_{L^2 \loc}$ denote a local $L^2$ norm, for example
the $L^2$-norm in a ball with large radius. One expects that this
difference actually approaches zero in local $L^2$ norm, i.e.,
\begin{equation}\label{as}
\lim_{t \to \infty} \inf_{ \Theta, E} \norm{ \psi(t) -Q_E \, e^{i
\Theta} }_{L^2 \loc} = 0 ~.
\end{equation}
If $- \Delta + V$ has only one bound state, it is proved in
\cite{SW1} \cite{PW} that the evolution will eventually settle
down to some ground state $Q_{E_\infty}$ with $E_\infty$ close to
$E$. Suppose now that  $- \Delta + V$ has two bound states: a
ground state $\phi_0$ with eigenvalue $e_0$ and an excited state
$\phi_1$ with eigenvalue $e_1$. It is proved in \cite{TY} that the
evolution with initial data $\psi_0$ near some $Q_E$ will
eventually settle down to some ground state $Q_{E_\infty}$ with
$E_\infty$ close to $E$.  See also \cite{BP} for the one
dimensional case, \cite{SW2} for nonlinear Klein-Gorden equations
with one unstable bound state.

If the initial data is not restricted to near the ground
states, the problem becomes much more delicate due to the presence of
the excited states. On physical ground, quantum mechanics tells us
that excited states are unstable and all perturbations should
result in a release of radiation and the relaxation of the excited
states to the ground states. Since bound states are periodic orbits,
this picture differs from the classical one
where periodic orbits are in general stable.

There were extensive linear analysis for bound states
of nonlinear Schr\"odinger and wave equations, see, e.g.,
 \cite{S, SS2, SS, G0, G,W1, W2}.
A special case of Theorem 3.5 of \cite{G}, page 330,  states that

\medskip

\noindent {\bf Theorem A} \ {\em Let $H_1 = - \Delta + V -E_1$.
The matrix operator
\[
JH_1 = \begin{bmatrix}0 & H_1\\- H_1&0\end{bmatrix}, \qquad J =
\begin{bmatrix}0 & 1\\- 1&0\end{bmatrix},
\]
is structurally stable if and only $e_0> 2e_1$. }

\medskip

The precise meaning of structural stability was given in \cite{G}.
Roughly speaking, it means that the operator remains stable under
small perturbations. Theorem A will not be directly used in this paper.

As we will see later, the linearized operator around an excited
state is a perturbation of $JH_1$. Thus, two different situations
occurs:

\medskip

\hspace*{1.5cm}1. Non-resonant case: $e_0 > 2 e_1$.
\quad ($e_{01}< |e_1|$).

\medskip

\hspace*{1.5cm}2. Resonant case: $e_0 < 2 e_1$.
\quad ($e_{01}> |e_1|$).

\medskip

\noindent Here $e_{01} = e_1 -e_0 > 0$. In the resonant case, Theorem A
says the linearized operator is in general unstable, which agrees with
the physical picture. In the non-resonant case, however, the linearized
operator becomes stable. The difference here is closely related to the
fact that $
2e_1-e_0$ lies in the continuum spectrum of $H_0$ only in the resonant
case.

In the resonant case, the unstable picture is confirmed for most
data near excited states in our work \cite{TY2}.   We prove that,
as long as the ground state component in $\psi_0 - Q_1$ is larger
than $\norm{\psi_0}^2$ times the size of the dispersive part
corresponding the continuous spectrum,
the solution will move away from the excited states
and relax and stabilize to ground states locally. Since
$\norm{\psi_0}^2$ is small, this assumption
allows the dispersive part to be  much larger than the ground state
component.

There is a small set of data where \cite{TY2} does not apply,
namely, those data with ground state component in $\psi_0 - Q_1$
smaller than $\norm{\psi_0}^2$ times the size of the dispersive
part. The aim of this paper is to show that this restriction
is almost optimal: we will construct within this small set
of initial data a ``hypersurface''  whose corresponding
solutions converge to
{\em excited states}.

This does not contradict with the physical intuition since this
hypersurface in certain sense has zero measure and
can not be observed  in experiments. These solutions, however, show
that linear instability does not imply all solutions to be unstable.
In the language of dynamical systems, {\it the  excited states are
one parameter family of hyperbolic fixed points and this
hypersurface is contained in the stable manifold of the fixed points}.
We believe that this surface is the whole stable manifold.

We will also construct solutions converging to excited states in
the non-resonant case, where it is expected since the linearized
operator is stable. We now state our assumptions on the potential $V$:

\noindent {\bf Assumption A0}: $H_0 := - \Delta + V$ acting on
$L^2(\R^3)$ has two simple eigenvalues $e_0<e_1< 0 $, with
normalized eigenvectors $\phi_0$ and $\phi_1$.

\noindent {\bf Assumption A1}:
The bottom of the continuous
spectrum to $-\Delta+ V$, $0$, is not a generalized
eigenvalue, i.e., not an eigenvalue nor a resonance.
There is  a small $\sigma>0$ such that
\[
|\nabla^\al V(x)| \le C \bkA{x}^{-5-\sigma}, \qquad \text{for }
|\al|\le 2 ~.
\]
Also, the functions $(x\cdot \nabla)^k V$, for $k=0,1,2,3$, are
$-\Delta$ bounded with a $-\Delta$-bound $<1$:
\begin{equation*}
  \norm{(x\cdot \nabla)^k V\phi}_2 \le \sigma_0 \norm{-\Delta\phi}_2 +
C\norm{\phi}_2,
  \qquad \sigma_0 < 1 , \quad k=0,1,2,3  ~.
\end{equation*}

Assumption A1 contains some standard conditions to assure that
most tools in linear Schr\"odinger operators apply. In particular,
it satisfies the assumptions of \cite{Y} so that the wave operator
$W_H=\lim_{t\to \infty} e^{i t H_0}e^{it\Delta}$
satisfies the $W^{k,p}$ estimates for $k\le 2$. These
conditions are certainly not optimal.

Let $e_{01}= e_1-e_0$ be the spectral gap of the ground state. In
the resonant case $2e_{01} > |e_0|$ so that $ 2e_1-e_0$ lies in
the continuum spectrum of $H_0$, we further assume

\noindent {\bf Assumption A2}: For some $s_0>0$,  
\begin{equation}
\label{A:gamma0} \gamma_0 \equiv \inf_{|s|<s_0} \lim_{\sigma \to
0+} \Im \bke{\phi_0 \phi_1^2, \, \frac 1 {H_0 + e_0 - 2 e_1 +s
-\sigma i} \Pc^{H_0}\phi_0 \phi_1^2} > 0 .
\end{equation}
Note that $\gamma_0 \ge 0$ since the expression above is
quadratic. This assumption is generically true.


Let $Q_1=Q_{1,E_1}$ be a nonlinear excited state with
$\norm{Q_{1,E_1}}_2$ small.
Since $(Q_1,E_1)$ satisfies \eqref{Q.eq},
the
function $\psi(t,x)=Q_1(x) e^{-iE_1t}$ is an exact solution of
\eqref{Sch}. If we consider solutions $\psi(t,x)$ of \eqref{Sch}
of the form
\[
\psi(t,x) = \bkt{ Q_1(x) + h(t,x)} \, e^{-i E_1 t}
\]
with $h(t,x)$ small in a suitable sense, then $h(t,x)$ satisfies
\[
\pd_t h = \L_1 h + \text{nonlinear terms}
\]
where $\L_1$, the {\it linearized operator around the
nonlinear excited state solution} $Q_1(x) e^{-iE_1t}$, is defined
by
\begin{equation} \label{L1.def}
\L_1 h = -i \bket{
(-\Delta + V -E_1 + 2 \la Q_1^2)\,h + \la Q_1^2 \,\wbar h\, }.
\end{equation}

\begin{theorem}  \label{th:1-2}
Suppose $H_0 = - \Delta + V$ satisfies assumptions A0--A1. Suppose
either

(NR) $e_0> 2e_1$, or

(R) $e_0< 2e_1$, and the assumption A2 for $\gamma_0$ holds.

\noindent Then there are $n_0>0$ and $\e_0(n)>0$ defined for $n
\in (0,n_0]$ such that the following holds. Let $Q_1: =Q_{1,E_1}$
be a nonlinear excited state with $\norm{Q_1}_{L^2}=n\le n_0$, and
let $\L_1$ be the corresponding linearized operator. For any $\xii
\in \Hc(\L_1)\cap (W^{2,1}\cap H^2)(\R^3)$ with
$\norm{\xii}_{W^{2,1}\cap H^2}=\e$, $0 < \e \le \e_0(n)$, there is
a solution $\psi(t,x)$ of \eqref{Sch} and a real function $\theta(t)
= O(t^{-1})$ for $t>0$ so that
\[
\norm{\psi(t) - \psias (t) }_{H^2} \le C \e^2  (1+t)^{-7/4},
\]
where $C=C(n)$ and
\[
\psias (t) =  Q_1 \, e^{-i E_1 t + i \theta(t)} + e^{-i E_1 t }e^{t\L_1}\xi_\infty .
\]

\end{theorem}

To prove this theorem, a detailed spectral analysis of the
linearized operator $\L_1$ is required. We shall classify the
spectrum of $\L_1$ completely in both non-resonant and resonant
cases, see Theorems \ref{th:2-1} and \ref{th:2-2}.
It is well-known that the continuous spectrum
$\Sigma_c$ of $\L_1$ is the same as that of $JH_1$, i.e.,
$\Sigma_c = \bket{si: s \in \R, |s| \ge |E_1|}$. The point spectrum
of $\L_1$ is more subtle. By definition, $H_1
\phi_1 = -(E_1-e_1)\phi_1$ and $H_1 \phi_0 = -(E_1-e_0) \phi_0$, and
thus the matrix operator $JH_1$ has 4 eigenvalues $\pm i(E_1-e_1)$ and
$\pm i(E_1-e_0)$.
In the non-resonant case, the eigenvalues of $\L_1$ are purely imaginary
and are small perturbations
of these eigenvalues. In the resonant case, the eigenvalues
$\pm i(E_1-e_0)$ are embedded inside the
continuum spectrum $\Sigma_c$. In general perturbation theory for
embedded eigenvalues, they turn into resonances under self-adjoint
perturbations. The operator $\L_1$ is however not a self-adjoint
perturbation of $H_1$. In this case, we shall prove that
{\it the embedded eigenvalues $\pm i(E_1-e_0)$ split into four eigenvalues
$\pm \evalue$ and $\pm \bar \evalue$ with the real part
given approximately by the  Fermi golden rule} (see \cite{RS}
Chap.XII.6):
\[
n^4 \Im \bke{\la \phi_0 \phi_1^2, \, \frac 1{-\Delta + V +e_0 -2
e_1-0i} \,\Pc \la \phi_1^2 \phi_0 } .
\]
Here $n$ is the size of $Q_1$, see \eqref{gamma.main}. In particular,
$e^{t \L_1}$ is {\em exponentially unstable} with the decay rate
(or the blow-up rate) given approximately by the  Fermi golden rule.
In other words, {\it although self-adjoint perturbation turns embedded
eigenvalues into resonances, the non-self adjoint perturbations
given by $\L_1$ turns an embedded eigenvalue into two eigenvalues
with the shifts in the real axis given to the leading order
by the  Fermi golden rule}. The dynamics of self-adjoint perturbation
of embedded eigenvalues were studied in  \cite{SW3}.

In the appendix we will  prove the existence of
solutions vanishing locally as $t \to
\infty$,  independent of the number of bound states in $H_0$.
Although some weaker versions of this proposition are expected,
it has never been proved in current form
and we include it for completeness.

\begin{proposition}  \label{th:1-1}
Suppose $H_0 = - \Delta + V$ satisfies assumption A1. There is an
$\e_0>0$ such that the following holds. For any $\xii \in
\Hc(H_0)\cap (W^{2,1}\cap H^2)(\R^3)$ with
$0<\norm{\xii}_{W^{2,1}\cap H^2}=\e \le \e_0$, there is a solution
$\psi(t,x)$ of \eqref{Sch} of the form
\[
\psi(t) = e^{-i tH_0} \xii + g(t), \qquad (t\ge 0),
\]
with $\norm{g(t)}_{H^2} \le C \e^2 (1+t)^{-2}$.
\end{proposition}

\section{Linear analysis for excited states}

As mentioned in \S 1, there is a family $\bket{Q_{1,E_1}}_{E_1}$ of
nonlinear excited states with the frequency $E_1$ as the parameter.
They satisfy
\begin{equation}   \label{Q1.eq}
    (-\Delta + V) Q_1 + \la |Q_1|^2 Q_1 = E_1Q_1  ~.
\end{equation}
Let $Q_1=Q_{1,{E_1}}$ be a fixed nonlinear excited state with
$n=\norm{Q_{1,{E_1}}}_2\le n_0 \ll 1$.
The linearized operator around the
nonlinear bound state solution $Q_1(x) e^{-iE_1t}$ is defined
in \eqref{L1.def}
\[
\L_1 h = -i \bket{
(-\Delta + V -E_1 + 2 \la Q_1^2)\,h + \la Q_1^2 \,\wbar h\, }.
\]
We will study the spectral properties of $\L_1$ in this section.
It properties are best understood in the complexification of
$L^2(\R^3,\Complex)$.

\begin{definition} \label{CL2}

Identify $\Complex$ with $\R^2$ and $L^2=L^2(\R^3, \Complex)$ with
$L^2(\R^3, \R^2)$. Denote by $\Complex L^2=L^2(\R^3, \Complex^2)$
the complexification of $L^2(\R^3, \R^2)$. $\Complex L^2$ consists
of 2-dimensional vectors whose components are in $L^2$. We have
the natural embedding
\[
f \in L^2 \longrightarrow  \vect{ \Re f \\ \Im f} \in \Complex
L^2.
\]
We equip $\Complex L^2$ with the natural inner product: For $f,g
\in \Complex L^2$, $f=\svect{f_1\\f_2}$, $g=\svect{g_1\\g_2}$, we
define
\begin{equation} \label{innerproduct}
(f,g) = \int \bar f \cdot g \,dx = \int (\bar f_1 g_1 + \bar f_2 g_2)\,dx .
\end{equation}
%
Denote by $\RE$ the operator first taking the real
part of functions in $\Complex L^2$ and then pulling back to
$L^2$:
\[
\RE: \Complex L^2 \to L^2, \quad \RE \vect{f\\g} = (\Re f)+i(\Re
g).
\]
\end{definition}

Recall the matrix operator $JH_1$ defined in Theorem A. Since $H_1
\phi_1 = -(E_1-e_1)\phi_1$ and $H_1 \phi_0 = -(E_1-e_0) \phi_0$,
the matrix operator $JH_1$ has 4 eigenvalues $\pm i(E_1-e_1)$ and
$\pm i(E_1-e_0)$ with corresponding eigenvectors
\begin{equation} \label{2-2A}
\vect{\phi_1\\- i \phi_1}, \quad \vect{\phi_1\\ i \phi_1}, \quad
\vect{\phi_0\\- i \phi_0}, \quad \vect{\phi_0\\ i \phi_0} .
\end{equation}
Notice that
\begin{equation} \label{2-3}
E_1-e_1 = O(n^2), \qquad E_1-e_0 = e_{01} + O(n^2).
\end{equation}
 The continuous spectrum of $JH_1$ is
\begin{equation}
\Sigma_c=\bket{ s i: \, s \in \R, |s| \ge |E_1|},
\end{equation}
which consists of two rays on the imaginary axis.

The operator $\L_1$ in its matrix form
\begin{equation} \label{3-3}
\begin{bmatrix}0 & L_-\\
 -L_+ &0\end{bmatrix}
,\qquad \text{with }
\left\{
\begin{aligned}
L_- &= - \Delta + V -E_1 + \la Q_1^2\\
L_+ &= - \Delta + V -E_1 + 3\la Q_1^2
\end{aligned} \right.
\end{equation}
is a perturbation of $JH_1$. By Weyl's lemma, the continuous
spectrum of $\L_1$ is also $\Sigma_c$. The eigenvalues are more
complicated. In both cases ($e_{01}<|e_1|$ and $e_{01}>|e_1|$)
they are near $0$ and $\pm i e_{01}$. As we shall see, in both
cases $0$ is an eigenvalue of $\L_1$. The main difference between
the two cases are the eigenvalues near $ ie_{01}$ and $-i e_{01}$.
If $e_{01}<|e_1|$, then $ ie_{01}$ lies outside the continuous
spectrum and $\L_1$ has an eigenvalue near $ie_{01}$ which is
purely imaginary.   On the other hand, if $e_{01}>|e_1|$, then $
ie_{01}$ lies inside the continuous spectrum. Generically it
splits under perturbation and the eigenvalues of $\L_1$ near $\pm
i e_{01}$ have non-zero real parts.

We shall show that $L^2(\R^3,\Complex)$, as a real vector space,
can be decomposed as the direct sum of three invariant subspaces
\begin{equation} \label{L2.dec}
L^2(\R^3, \Complex) = S(\L_1)\oplus \eigen(\L_1) \oplus \Hc(\L_1)
\end{equation}
Here $S(\L_1)$ is the generalized null space, $\eigen(\L_1)$ is a
generalized eigenspaces and $\Hc(\L_1)$ corresponds to the
continuous spectrum. Both  $S(\L_1)$ and $\eigen(\L_1)$ are finite
dimensional.

Recall the Pauli matrices
\[
\sigma_1 = \begin{bmatrix}0&1\\1&0\end{bmatrix},\quad \sigma_2 =
\begin{bmatrix}0&-i\\i&0\end{bmatrix},\quad \sigma_3 =
\begin{bmatrix}1&0\\0&-1\end{bmatrix}.
\]
They are self-adjoint and
\begin{equation}\sigma_1 \L_1 =\L_1^*
\sigma_1, \qquad \sigma_3 \L_1 = - \L_1\sigma_3 ,
\end{equation}
where $\L_1^* =\bkt{\begin{smallmatrix}0& -
L_+\\L_-&0\end{smallmatrix}}$.

Let $R_1 = \pd_{E_1} Q_{1,E_1}$. Direct
differentiation of \eqref{Q1.eq} with respect to $E_1$ gives  $L_+
R_1 = Q_1$. Since $L_-Q_1=0$ and $L_+ R_1 = Q_1$, we have $\L_1
\svect{0\\Q_1} = 0$ and $\L_1 \svect{R_1\\0} = -\svect{0\\Q_1}$.
We will show dim$_\R S(\L_1)=2$, hence
\begin{equation} \label{S.def} S(\L_1)= \myspan_\R \bket{\vect{0\\ Q_1},
\vect{R_1\\0}} . \end{equation}

$\Hc(\L_1)$ can be characterized as
\begin{equation} \label{Hc.def}
\Hc(\L_1) = \bket{ \psi \in \Complex L^2: (\sigma_1 \psi, f)=0 , \
\forall f \in S(\L_1)\oplus \eigen(\L_1) }.
\end{equation}
We will use \eqref{Hc.def} as a working definition of $\Hc(\L_1)$.
After we have proved the spectrum of $\L_1$ and the resolvent
estimates, we will use the wave operator of $\L_1$ (see
\cite{C,Y,Y2}) to show that \eqref{Hc.def} agrees with the usual
definition of the continuous spectrum subspace. See \S 2.5.

The space $\eigen(\L_1)$, however, has very different properties
in the two cases, due to whether $\pm i (E_1-e_0)$ are embedded
eigenvalues of $JH_1$. We will consider $\eigen = \eigen(\L_1)$ as
a subspace of $L^2(\R^3, \R^2)$ and denote by $\Complex \eigen
\subset \Complex L^2$ the complexification of $\eigen$. We will
show that $\Complex \eigen$ is a direct sum of eigenspaces of
$\L_1$ in $\Complex L^2$. We also have
\begin{equation} \label{SEorth}
(\sigma_1 f,g)=0 , \quad  \forall f \in S(\L_1), \ \forall g \in
\eigen(\L_1).
\end{equation}
We have the following two theorems for the two cases.

{\bf Remark} The case $e_0  = 2e_1$: The spectral property of
$\L_1$ is not clear.

\begin{theorem} [Non-resonant case] \label{th:2-1}
Suppose  $e_0> 2e_1$, and the assumptions A0-A1 hold. Let
$Q_1=Q_{1,E_1}$ be a nonlinear excited state with sufficiently
small $L^2$-norm, and let $\L_1$ be defined as in \eqref{L1.def}.

(1) The eigenvalues of $\L_1 $ are  $0$ and $\pm \evalue $. The
multiplicity of $0$ is two. The other eigenvalues are simple.
Here $\evalue =i\ka$, $\ka$ is real, $\ka = e_{01} + O(n^2)$.
There is no embedded eigenvalue. The bottoms of the continuous
spectrum are not eigenvalue nor resonance.

(2) The space $L^2= L^2(\R^3,\Complex)$, as a real vector space,
can be decomposed as in \eqref{L2.dec}. Here $ S(\L_1)$ and
$\Hc(\L_1)$ are given in \eqref{S.def} and \eqref{Hc.def},
respectively; $\eigen(\L_1)$ is the space corresponding to the
perturbation of the eigenvalues $\pm i (E_1-e_0)$ of $JH_1$. We
have the orthogonality relation \eqref{SEorth}.

(3) Let $ \Complex \eigen$ denotes the complexification of
$\eigen=\eigen(\L_1)$. $ \Complex \eigen$ is
2-complex-dimensional. $\eigen$ is 2-real-dimensional. We have
\begin{equation} \label{E1.def}
\begin{split}
\Complex \eigen &= \myspan_\Complex
\bket{\evector,\wbar \evector}, \\
\eigen &= \myspan_\R \bket{\svect{u\\0},\svect{0\\v} }.
\end{split}
\end{equation}
Here $\evector = \svect{u\\ -i v}$ is an eigenfunction of $\L_1$
with eigenvalue $\evalue $. $u$ and $v$ are real-valued
$L^2$-functions satisfying $L_+ u =-\ka v$, $L_- v =-\ka u$ and
$(u,v)=1$. $u$ and $v$ are perturbations of $\phi_0$. $\bar
\evector= \svect{u\\ i v}$ is another eigenfunction with
eigenvalue $- \evalue $. We have $\L_1 \evector = \evalue
\evector$, $\L_1 \bar \evector = - \evalue \bar \evector$.

(4) For any function $\zeta \in \eigen(\L_1)$, there is a unique
$\al \in \Complex$ so that
\[
\zeta = \RE \al \evector,
\]
and we have $\L_1 \zeta =  \RE \evalue  \al \evector $,
$e^{t\L_1} \zeta = \RE e^{t\evalue } \al \evector $.

(5) We have the orthogonality relations in \eqref{Hc.def} and
\eqref{SEorth}. Hence any $\psi \in L^2$ can be decomposed as (see
\eqref{L2.dec})
\begin{equation} \label{L2.dec2} \psi = a\vect{R_1 \\0} +b
\vect{0\\ Q_1} + c\vect{ u \\ 0} + d \vect{0\\v} + \eta
\end{equation}
with $\eta \in \Hc(\L_1)$,
\begin{equation} \label{3-10}
\begin{gathered}
a=(Q_1,R_1)^{-1}(Q_1,\Re \psi),\\
b=(Q_1,R_1)^{-1}(R_1,\Im \psi),
\end{gathered}
\qquad
\begin{gathered}
c=(u,v)^{-1}(v,\Re \psi),\\
d=(u,v)^{-1}(u,\Im \psi).
\end{gathered}
\end{equation}

(6) Let $M_{1}\equiv \eigen(\L_1 ) \oplus \Hc(\L_1 )$. We have
\begin{equation}
M_{1}\equiv \eigen(\L_1 ) \oplus \Hc(\L_1 )
 = \begin{bmatrix} Q_1^\perp \\ R_1^\perp \end{bmatrix}.
\end{equation}
There is a constant $C_2>1$ such that,
for all $\phi \in M_{1}$ and all $t\in \R$, we have
\begin{equation} \label{3-12}
C_2^{-1}\norm{\phi}_{H^k} \le  \norm{e^{t\L_1}\phi}_{H^k} \le C_2
\norm{\phi}_{H^k} ~, \qquad (k=1,2).
\end{equation}

(7) Decay estimates: For all $\eta \in \Hc(\L_1)$,
for all $p \in [2,\infty]$, one has
\[
\norm{e^{t\L_1}\eta}_{L^p} \le C |t|^{- 3(\frac 12 - \frac 1p)}
\norm{\eta}_{L^{p'}}.
\]

\end{theorem}

\begin{theorem} [Resonant case] \label{th:2-2}
 Suppose  $e_0< 2e_1$, and the assumptions A0-A2 hold. Let
$Q_1=Q_{1,E_1}$ be a nonlinear excited state with sufficiently
small $L^2$-norm, and let $\L_1$ be defined as in \eqref{L1.def}.

(1) The eigenvalues of $\L_1$ are $0$,  $\pm \evalue $ and $\pm
\bar \evalue $. The multiplicity of $0$ is two. The other eigenvalues
are simple. Here $\evalue = i \ka +
\gamma$, $\ka, \gamma >0 $, $\ka = e_{01} + O(n^2)$, and $ \frac
34 \la^2 \gamma_0 n^4\le \gamma \le C n^4$. ($\gamma_0$ is given
in \eqref{A:gamma0}). There is no embedded eigenvalue. The bottoms
of the continuous spectrum are not eigenvalue nor resonance.

There is an $\evalue$-eigenvector $\evector$, $\L_1 \evector =
\evalue \evector$, which is of order one in $L^2$  and $\evector-
\svect{\phi_0\\-i\phi_0}$ is locally small in the sense that
\begin{equation}
\label{3-18} \length{(\phi , \evector- \svect{\phi_0\\-i\phi_0})}
\le C  n^2 \norm{\bkA{x}^{r} \phi}_{L^2},
\end{equation}
for any $r>3$. However, $\evector$ is not a perturbation of
$\svect{\phi_0\\-i\phi_0}$ in $\Complex L^2$. In fact, $\evector =
\svect{u \\ v}$ with $u- \phi_0$ and $v+ i\phi_0$ of order one in
$L^2$,
\[
u =\phi_0 - \frac { 1}{- \Delta  + V- E_1  - \ka + \gamma i} \,
\Pc(H_0) \la \phi_0 Q_1^2 + O(n^2) \quad \text{in }L^2,
\]
and $v = - L_+ u/\evalue$.
Note  $ -E_1 -  \ka = e_0 - 2e_1 + O(n^2)$.

 (2) The space
$L^2= L^2(\R^3,\Complex)$, as a real vector space, can be
decomposed as in \eqref{L2.dec}. Here $ S(\L_1)$ and $\Hc(\L_1)$
are given in \eqref{S.def} and \eqref{Hc.def}, respectively;
$\eigen(\L_1)$ is the space corresponding to the perturbation of
the eigenvalues $\pm i (E_1-e_0)$ of $JH_1$. We have the
orthogonality relation \eqref{SEorth}.

(3) Let $ \Complex \eigen$ denotes the complexification of
$\eigen=\eigen(\L_1)$.  $ \Complex \eigen$ is
4-complex-dimensional.  $\eigen$ is 4-real-dimensional. If we
write $\evector=\svect{u\\ v}= \vect{u_1+u_2i\\
v_1+ v_2 i}$ with $u_1,u_2, v_1, v_2$ real-valued $L^2$ functions,
we have
\begin{equation} \label{3-16}
\begin{aligned} \Complex \eigen&= \myspan_\Complex
\bket{\evector,\bar \evector, \sigma_3 \evector, \sigma_3 \bar \evector},\\
\eigen&= \myspan_\R
\bket{\vect{u_1\\0},\vect{u_2 \\0},\vect{0\\ v_1},\vect{0\\ v_2}}.
\end{aligned}
\end{equation}
Recall $\sigma_3 = \bkt{
\begin{smallmatrix}1&0\\ 0&-1\end{smallmatrix}}$. The other
eigenvectors are $\bar \evector$, $\sigma_3 \evector$ and
$\sigma_3 \bar \evector$,
\begin{equation} \label{3-17}
\L_1 \evector= \evalue  \evector, \quad
\L_1 \bar \evector= \bar\evalue  \bar\evector, \quad
\L_1 \sigma_3\evector= -\evalue  (\sigma_3\evector), \quad
\L_1  \sigma_3\bar\evector= -\bar \evalue  (\sigma_3\bar \evector).
\end{equation}

(4) For any function $\zeta \in \eigen(\L_1)$, there is a unique
pair $(\al,\beta) \in \Complex^2$ so that
\begin{equation} \label{zeta.dec2}
\zeta = {\bf RE} \bket{\al \evector + \beta \sigma_3 \evector},
\end{equation}
and we have $\L_1 \zeta =  \RE \bket{\evalue  \al \evector -
\evalue  \beta \sigma_3 \evector} $, $e^{t\L_1} \zeta = \RE
\bket{e^{t\evalue } \al \evector +e^{-t\evalue } \beta \sigma_3
\evector} $.

(5) We have the orthogonality relations in \eqref{Hc.def} and
\eqref{SEorth}.  Moreover, $\sigma_1 \bar \evector \perp
\bket{\bar \evector, \sigma_3 \evector, \sigma_3 \bar \evector}$,
$\sigma_1 \evector \perp  \bket{\evector, \sigma_3 \evector,
\sigma_3 \bar \evector}$, and $\int \bar u v dx =0$, etc.
For any function $\psi \in \Complex L^2$, if we
decompose
\begin{equation}
\psi = a \vect{R_1\\0} + b \vect{0\\Q_1} + \al_1
\evector + \al_2 \bar \evector + \beta _1 \sigma_3 \evector +
\beta_2 \sigma_3 \bar \evector + \eta
\end {equation}
where $a,b,\al_1,\al_2,\beta_1,\beta_2 \in \Complex$ and $\eta \in
\Hc(\L_1)$, then we have
\begin{gather}
a= c_1 (\sigma_1 \svect{0\\Q_1},\psi),  \qquad
b= c_1 (\sigma_1 \svect{R_1\\0},\psi), \nonumber
\\
\al_1 = c_2 (\sigma_1 \bar \evector,\psi),  \qquad
\al_2 = \bar c_2 (\sigma_1  \evector,\psi), \label{3-21}
\\
\beta_1 = -c_2 (\sigma_1 \sigma_3\bar \evector,\psi),  \qquad
\beta_2 = - \bar c_2 (\sigma_1  \sigma_3\evector,\psi),  \nonumber
\end{gather}
where  $c_1^{-1} = (Q_1,R_1)$ and
$c_2^{-1}= (\sigma_1 \bar \evector, \evector) = \int 2 u v dx$.
(Note $c_1 \la >0$.)
The statement that $\psi \in \eigen$ is equivalent to that $a,b
\in \R$, $\al_1 =\al_2=\al/2$, $\beta_1 = \beta_2=\beta/2$ and
$\eta$ is real. In this case,
\begin{equation}
\psi = a \vect{R_1\\0} + b \vect{0\\Q_1}
+ {\bf RE} \bket{\al \evector + \beta \sigma_3 \evector} + \eta,
\end{equation}
with $a,b\in \R$, $\eta \in \Hc(\L_1)$ real, $\al, \beta \in \Complex$,
and
\begin{equation} \label{PalPbeta.def}
\al = P_\al(\psi) \equiv 2 c_2 (\sigma_1 \bar \evector,\psi),  \quad
\beta = P_\beta (\psi)\equiv
-2 c_2 (\sigma_1 \sigma_3\bar \evector,\psi).
\end{equation}
$P_\al$ and $P_\beta$ are maps from $L^2$ to $\Complex$.

(6) There is a constant $C>1$ such that,
for all $\eta \in \Hc(\L_{1})$ and all $t\in \R$, we have
\[
C^{-1}\norm{\eta}_{H^k} \le  \norm{e^{t\L_1}\eta}_{H^k} \le C
\norm{\eta}_{H^k} ~, \qquad (k=1,2).
\]

(7) Decay estimates: For all $\eta \in \Hc(\L_1)$,
for all $p \in [2,\infty]$, one has
\[
\norm{e^{t\L_1}\eta}_{L^p} \le C |t|^{- 3(\frac 12 - \frac 1p)}
\norm{\eta}_{L^{p'}},
\]
where $C=C(n,p)$ depends on $n$.

\end{theorem}

\myremark (i). In (6), we restrict ourselves to $\Hc(\L_{1})$, not
$M_1$ as in Theorem \ref{th:2-1}. (ii). In (3), $\evector$ is not
a perturbaiton of $\svect{\phi_0\\-i \phi_0}$. Also, the $L^2$
functions $u_1$ and $u_2$ are independent of each other. So are
$v_1$ and $v_2$. (iii) In (7) the constant depends on $n$ since
there are eigenvalues which are very close to the continuous
spectrum.

Since the proof of Theorem \ref{th:2-1} is easier, we postpone it
to the last subsection \S \ref{sec:pf21}.  We will focus on
proving Theorem \ref{th:2-2} in the following subsections.

\subsection{Perturbation of embedded eigenvalues and their eigenvectors}

In this subsection we study the eigenvalues of $\L_1$ near
$ie_{01}$. By symmetry we get also the information near $-i
e_{01}$.

For our fixed nonlinear excited state $Q_1=Q_{1,E_1}$, let
$H=-\Delta + V - E_1 + \la Q_1^2$. ($H$ is $L_1$ in \eqref{3-3}.)
Let $ \gs$ denote a positive normalized ground state of $H$, with
ground state energy $-\rho$ which is very close to $-e_{01}$.
Hence the bottom of the continuous spectrum of $H$, which is close
to $|e_1|$, is less than $\rho$. We have
\[
HQ_1=0, \qquad H\gs = -\rho \gs.
\]
\begin{equation} \label{5-1A}
Q_1 = n \phi_1 + O(n^3), \qquad \gs = \phi_0 +
O(n^2).
\end{equation}

We want to solve the eigenvalue problem $\L_1 \evector = \evalue \evector$
with $\evalue $ near $i e_{01}$. Write $\evector = \svect{u\\v}$. The
problem has the form
\[
\begin{bmatrix}0& H \\ -(H + 2\la Q_1^2) & 0 \end{bmatrix}
\vect{u \\ v} =  \evalue \vect{u \\ v}
\]
for some $\evalue$ near $i e_{01}$ and for some complex
$L^2$-functions $u,v$. We have
\[
H v =  \evalue u, \qquad  (H+2\la Q_1^2) u = - \evalue v.
\]
Thus $H (H+2\la Q_1^2) u = - \evalue^2 u$. Suppose $\evalue= i \ka
+ \gamma $ with $\gamma>0$. Since $\Im (- \evalue^2)<0$ and $H$ is
real, it is more convenient to solve
 $H (H+2\la Q_1^2) \bar u = - \bar \evalue^2 \bar u$ instead.
If we decompose $\bar u= a
\gs + b Q_1 + h$ with $h\in \Hc(H)$, we find $b=0$ since $\bar u \in
{\rm Image }\, H$. Since $a \not = 0$, we may assume $\bar u = \gs +
h$. Let $A= H 2\la Q_1^2$ and $z= - \bar \evalue^2 \sim  \rho^2 $.
(A small $\Re \evalue >0$ corresponds to a small $\Im z > 0$.)
We have
\[
(H^2 + A) (\gs + h) = z (\gs + h),
\]
i.e.,
\begin{equation} \label{A-1}
 z \gs + z h =\rho^2\gs + A \gs + (H^2 + A) h.
\end{equation}

Taking projection $\Pc=\Pc(H)$, we get
\[
z h=\Pc A \gs + (H^2 + \Pc A \Pc) h ~.
\]
If $\Im z \not = 0$,
\begin{equation} \label{A-1.5}
h = -(H^2 + \Pc A \Pc -z)^{-1}\Pc A \gs .
\end{equation}
On the other hand, if $\Im z = 0$, then $h$ is generically not
in $L^2$. We will assume $\Im z \not = 0$ in this subsection.
The non-existence of eigenvalues with $\Im z  = 0$ will be proved
in next subsection.

Taking inner product of \eqref{A-1} with $ \gs$, we get
\[
z = \rho^2 + (\gs, A \gs) + \bke{\gs ,  A h}  .
\]
Substituting \eqref{A-1.5}, we get
\begin{equation} \label{A-2}
z= \rho^2+ (\gs, A \gs) -
\bke{\gs , A(H^2 + \Pc A \Pc -z)^{-1}\Pc A \gs } .
\end{equation}

If $A$ is self-adjoint, then the signs of the imaginary parts of
the two sides of the equation are different. Thus $z$ is real and
generically $h$ is not in $L^2$. In our case, $A= H 2 \la Q_1^2$
is not self-adjoint. Recall $H \gs = - \rho \gs$. Equation
\eqref{A-2} becomes the following fixed point problem
\begin{equation}
\label{5-5} z = f(z)
\end{equation}
where
\begin{multline}
f(z) = \rho^2 -\rho (\gs 2\la Q_1^2 \gs)
\\
+ \rho \bke{\gs 2 \la Q_1^2, \,
(H^2 + H \Pc 2 \la Q_1^2 \Pc -z)^{-1}H\Pc 2 \la Q_1^2 \gs }.
\label{f.def}
\end{multline}

Let
\begin{equation} \label{3-7}
R(z) = (H^2 -z)^{-1}H = \frac 1{ 2(H - \sqrt z)} + \frac 1{ 2(H + \sqrt z)},
\end{equation}
where $\sqrt z$ takes the branch $\Im \sqrt z >0$ if $\Im z > 0$.
We can expand $f(z)$ as
\begin{equation} \label{3-8}
f(z) =  \rho^2 -\rho(\gs 2\la Q_1^2 \gs) +\sum_{k=1}^\infty \rho 2 \la
\bke{\gs Q_1, \bkt{2 \la Q_1\Pc R(z) \Pc Q_1}^k Q_1 \gs }.
\end{equation}

Let
\begin{align*}
z_0 &= \rho^2 -\rho(\gs 2 \la Q_1^2 \gs),
\\
z_1 &= z_0 +  4 \rho \la^2\bke{\gs Q_1^2, \, R(z_0 + 0i)\Pc Q_1^2 \gs }.
\end{align*}
We have $|z_1 - z_0|\le C n^4$ from its explicit form,
(cf.~\eqref{311-1} of Lemma \ref{th:A-2} below). We also have, by
\eqref{5-1A} and \eqref{A:gamma0}
\[
\Im z_1 = \Im 4\rho \la^2\bke{\gs Q_1^2, \,
\frac 1{2(H - 0i)} \,\Pc Q_1^2 \gs }
\ge  \tfrac 74 \, e_{01} \la^2  n^4 \, \gamma_0 + O(n^6) > 0 .
\]

Let $r_0 = \frac 14( (e_{01})^2 - |e_1|^2)$ be a length of order
1. Denote the regions
\begin{equation} \label{G.def}
G = \bket{x+iy:|x-\rho^2|< r_0, \ 0<y<r_0},
\end{equation}
\begin{equation} \label{D.def}
D = B(z_1, n^5) = \bket{z: |z-z_1| \le n^5\ }.
\end{equation}
Clearly $z_0 \in G$, $z_1 \in D \subset G$. Also observe that the
real part of all points in $G$ are greater than $|E_1|^2$.
We will solve the fixed point problem \eqref{5-5} in $D$. We need
the following two lemmas.

\begin{lemma} \label{th:A-2}
Fix $r>3$. There is a constant $C_1>0$ such that, for all $z \in G$,
\begin{equation} \label{311-1}
\norm{\bkA{x}^{-r} \Pc R(z) \Pc \bkA{x}^{-r}}_{(L^2,L^2)}\le C_1 ,
\end{equation}
\begin{equation} \label{311-2}
\norm{\bkA{x}^{-r} \Pc \frac {d}{dz} R(z) \Pc \bkA{x}^{-r}}_{(L^2,L^2)}
\le C_1 (\Im z)^{-1/2}.
\end{equation}
Here $\Pc = \Pc(H)$.
Moreover, for $w_1,w_2\in G$, 
\begin{multline} \label{311-3}
\norm{\bkA{x}^{-r} \Pc [R(w_1)-R(w_2)] \Pc \bkA{x}^{-r}}_{(L^2,L^2)}
\\
\le C_1 (\max (\Im w_1, \Im w_2))^{-1/2}|w_1-w_2|.
\end{multline}

\end{lemma}

\myproof
We have
\begin{equation}
R(z) = (H^2 -z)^{-1}H = \frac 1{ 2(H - \sqrt z)} + \frac 1{ 2(H + \sqrt z)}
\end{equation}
Since  $\frac 1{ 2(H + \sqrt z)}$ is regular in a neighborhood of $\bar G$,
it is sufficient to prove the lemma with $R(z)$ replaced by
$R_1(z):=(H - \sqrt z)^{-1}$.

That $\norm{\bkA{x}^{-r} \Pc R_1(z) \Pc \bkA{x}^{-r}}
_{(L^2,L^2)}\le C_1$ is well-known, see e.g. \cite{Agmon},
\cite{JK}. The estimate \eqref{311-2} will follow from
\eqref{311-3} by taking limit. We now show \eqref{311-3} for
$R_1(z)$. For any $w_1,w_2 \in G$, we have $|\sqrt {w_1} -\sqrt
{w_2}| \le |w_1 -w_2|$. Write $\sqrt {w_1} = a_1 + i b_1 $ and
$\sqrt {w_2} = a_2 + i b_2 $.  We may assume $0 < b_1 < b_2$. Let
$w_3 \in G$ be the unique number such that $\sqrt {w_3} = a_1 + i
b_2 $.

For any $u,v \in L^2$ with $\norm{u}_2=\norm{v}_2=1$, let $u_1 =
\Pc \bkA{x}^{-r} u$, $v_1 = \Pc \bkA{x}^{-r} v$. We have $u_1,v_1
\in L^1\cap L^2(\R^3)$ and
\begin{align*}
&\length{\bke{u,\bkA{x}^{-r} \Pc \bkt{R_1(w_1)-R_1(w_3)} \Pc
\bkA{x}^{-r}v}}
\\
&=\length{\int_0^\infty (u_1, e^{-it(H- a_1)}v_1) \, (e^{-b_1 t} -
e^{-b_2 t}) \, dt}
\\
&\le \int_0^\infty (1+t)^{-3/2} (e^{-b_1 t} - e^{-b_2 t}) \, dt
\le C ( b_2^{1/2} - b_1^{1/2}) \le b_2^{-1/2} (b_2 - b_1).
\end{align*}
Here we have used the decay estimate for
$e^{-itH}$ with $H=-\Delta + V -E_1 - \la Q_1^2$, namely,
\begin{equation} \label{decayest}
\norm{ e^{-itH} \Pc \phi}_{L^\infty} \le C |t|^{- 3/2}
\norm{\phi}_{L^1}
\end{equation}
under our Assumption A1. See \cite{JK,JSS,Y}.

We also have
\begin{align*}
&\length{\bke{u,\bkA{x}^{-r} \Pc \bkt{R_1(w_3)-R_1(w_2)} \Pc \bkA{x}^{-r}v}}
\\
&=\length{\int_0^\infty (u_1, e^{-it(H- s_2 -i b_2)}v_1)  \,
(e^{i(a_1-a_2) t} - 1) \, dt}
\\
&\le \int_0^\infty (1+t)^{-3/2} e^{-b_2 t}\, |e^{i(a_1-a_2) t} -
1|\, dt \le C b_2^{-1/2} |a_1-a_2|.
\end{align*}
Since $|a_1-a_2| + |b_1-b_2| \sim |\sqrt {w_1} -\sqrt {w_2}| \le
|w_1 -w_2|$, we conclude
\[
\length{\bke{u,\bkA{x}^{-r} \Pc \bkt{R_1(w_1)-R_1(w_2)} \Pc
\bkA{x}^{-r}v}} \le C b_2^{-1/2} |w_1 - w_2|.
\]
Hence we have \eqref{311-3}.
\myendproof

\begin{lemma} \label{th:A-1}
Recall the regions $G$ and $D$ are defined in \eqref{G.def}--\eqref{D.def}.

(1) $f(z)$ defined by \eqref{f.def} is well-defined and analytic in $G$.

(2) $|f'(z)|\le C n^4 (\Im z)^{-1/2}$ in $G$ and $|f'(z)|\le 1/2$
in $D$.

(3) for $w_1,w_2 \in G$,
\[
|f(w_1) - f(w_2)|
\le C n^4 (\max (\Im w_1, \Im w_2))^{-1/2}|w_1-w_2|.
\]

(4) $f(z)$ maps $D$ into $D$.
\end{lemma}

\myproof By \eqref{311-1}, the expansion \eqref{3-8} can be
bounded by
\[
|f(z)| \le C + CC_1 n^4 + C C_1^2 n^6 + \cdots
\]
and thus converges. Since every term in \eqref{3-8} is analytic,
$f(z)$ is well-defined and analytic. We also get the estimates in
(2). To prove (3), let $b=\max (\Im w_1, \Im w_2)$. Then from
\eqref{311-1}--\eqref{311-3},
\[
|f(w_1)-f(w_2)|\le
\sum_{k=1}^\infty C kC_1^k n^{2k+2} b^{-1/2} |w_1-w_2| \le C n^4
b^{-1/2}|w_1-w_2|.
\]

It remains to show (4). We first estimate $|f(z_1)-z_1|$. Write
$z_1 = z_0 + a + bi$. Recall that $|a|<C n^4$ and $\frac 14 \la ^2
\gamma_0 n^4< |b| <C n^4$. Using \eqref{311-3} and \eqref{311-1}
we have
\begin{align*}
|f(z_1)-z_1| &= \bke{\gs Q_1^2, \bkt{R(z_1) - R(z_0 + 0i)} \Pc Q_1^2 \gs }
\\
&\quad + \sum_{k=2}^\infty
\bke{\gs Q_1, \bkt{Q_1\Pc R(z_1) \Pc Q_1}^k Q_1 \gs }
\\
&\le
C n^4 b^{-1/2} (|a|+|b|) 
+ CC_1^2 n^6 + C C_1^3 n^8 + \cdots \le   C n^6
\end{align*}
Hence $|f(z_1)-z_1| \le C n^6$.
For any $ z\in D$, we have
\[
|f(z)-z_1| \le |f(z) - f(z_1)| + | f(z_1) - z_1|
\le \frac 12 |z-z_1| +  C n^6 \le  |z-z_1|.
\]
Hence $f(z) \in D$. This proves (4). \myendproof

We are ready to solve \eqref{5-5} in $G$. By Lemma \ref{th:A-1}
(1), (2) and (4), the map $f \to f(z)$ is a contraction mapping in
$D$ and hence has a unique fixed point $z_*$ in $D$. By (3), for
any $z\in G$ we have $|f(z) - f(z_*)|\le C n^4 (\Im z_*)^{-1/2}
|z-z_*| \le \frac 12  |z-z_*|$. Hence there is no other fixed
point of $f(z)$ in $G$.

By symmetry, there is another unique fixed point with negative
imaginary part. Moreover, they have the size indicated in Theorem
\ref{th:2-2}. We will prove in Subsections 2.2 and 2.3 that
$\evalue$ does not admit generalized eigenvectors and that there
is no purely imaginary eigenvalue near $ie_{01}$, i.e., there is
no embedded eigenvalue. Hence $\evalue,$ and $- \bar \evalue$ are
simple and are the only eigenvalues near $ie_{01}$.

\bigskip

We now look more carefully on $z_*$ and $u_*$, where $u_*$ denotes
the unique solution of $H (H+2\la Q_1^2) u_* = - \evalue^2 u_*$
with the form $u_*= \gs + \bar h_*$.

Recall $|z_1 - z_*|\le n^5$ and
\[ z_1
= \rho^2 -\rho(\gs 2 \la Q_1^2 \gs) +  4 \rho \la^2\bke{\gs Q_1^2,
\, R(z_0 + 0i)\Pc Q_1^2 \gs },
\]
where $z_0=\rho^2 -\rho(\gs 2 \la Q_1^2 \gs)$. Hence
\begin{align*}
 \sqrt {z_*} &=  \sqrt {z_1} + O(n^5)
\\
&=  \rho  -(\gs  \la Q_1^2 \gs) +  2  \la^2\bke{\gs Q_1^2,
\, R(z_0 + 0i)\Pc Q_1^2 \gs }
\\
& \quad + \frac 1{4\rho} (\gs  \la Q_1^2 \gs) ^2 + O(n^5).
\end{align*}
Since $z_* = - \bar \evalue^2$, we have $\bar \evalue = i\sqrt {z_*}$.
Thus if we write $\evalue = i \ka + \gamma $, then
\begin{align}
\ka &= \rho  - (\gs  \la Q_1^2 \gs)
+ \frac 1{4\rho} (\gs  \la Q_1^2 \gs) ^2 \nonumber
\\
& \qquad +  \Re 2 \la^2  \bke{\gs Q_1^2, \, R(z_0 + 0i)\Pc Q_1^2 \gs }
+ O(n^5), \label{ka.main}
\\
\nonumber
\gamma &= -\Im 2 \la^2\bke{\gs Q_1^2, \, R(z_0 + 0i)\Pc Q_1^2 \gs }+ O(n^5).
\end{align}
By  \eqref{3-7}, \eqref{5-1A}, and expansion into series,
\begin{align}
\gamma &= \Im  \la^2\bke{\gs Q_1^2, \, (H-\sqrt{z_0} - 0i)\Pc Q_1^2 \gs }+ O(n^5)
\nonumber
\\
&=\Im \la^2 n^4\bke{\phi_0 \phi_1^2, \, \frac 1{-\Delta + V -
E_1-\sqrt{z_0}-0i} \,\Pc \phi_1^2 \phi_0 } + O(n^5).
\label{gamma.main}
\end{align}
By \eqref{A:gamma0}, $\gamma\ge \la^2  n^4 \, \gamma_0 + O(n^5)$ .

By \eqref{A-1.5} and $A=H2 \la Q_1^2$, we have
\begin{equation}
h_* =  -(H^2 + \Pc H2 \la Q_1^2 \Pc -z_*)^{-1}H \Pc 2 \la Q_1^2 \gs,
\end{equation}
where $\Pc = \Pc(H)$. We now expand the resolvent on the right
side as in \eqref{3-8}. Then by Lemma \ref{th:A-2}, we can derive
$|(\phi, h)|\le C n^2 \norm{\bkA{x}^r\phi}_2$, for any $r>3$.

We now show that $h_*$ is bounded in $L^2$ with a bound uniform in
$n$. Recall $\sqrt{z_*} = \ka + i \gamma$ with $\ka \sim e_{01}$,
$\gamma > \tfrac 12 \la^2 \gamma_0 n^4$. Since $Q_1 = n\phi_1 +
O(n^3)$, by expansion and \eqref{5-1A} we have
\begin{align}
h_* &=  - (H^2 -z_*)^{-1}H \Pc(H)  2 \la \phi_0 Q_1^2 + O(n^2)
\nonumber
\\
&=  - (H - \sqrt {z_*})^{-1} \Pc(H)     \la \phi_0 Q_1^2 + O(n^2)
\nonumber
\\
&=  - \frac { 1}{- \Delta + V + s - \gamma i} \Pc(H_0) \la \phi_0
Q_1^2 + O(n^2), \label{5-u.dec}
\end{align}
where $s=   - E_1  - \ka = e_0 - 2e_1 + O(n^2)$. Here we have used
the fact that
\[ \Pc(H) \phi= \Pc(H_0) \phi + n^2 \sum_{k=1}^N
(\psi_k^*,\phi) \psi_k
\]
for some local functions $\phi_k$, $\phi_k^*$ of order one. We
will show that the leading term on the right of \eqref{5-u.dec} is
of order one in $L^2$. It follows from the same proof that
$O(n^2)$ on the right is also in $L^2$ sense.

Observe that, for $f(p) \in L^2$ with $\norm{f}_2 \le 1$,
\begin{align*}
&\int \frac 1{p^2 - s + \gamma i} f(p) \cdot \frac 1{p^2 - s -
\gamma i} \bar f(p) \, dp
\\
&= \int |f(p)|^2 \frac 1{(p^2 - s)^2 + \gamma^2 } \, dp
\\
&\le C + C\int _{\sqrt s/2 }^{3\sqrt s/2} \frac 1{(|r-\sqrt s| +
\gamma)^2} \, dr
\\
&= C + 2C \int _0^{\sqrt s/2} \frac       1{(r + \gamma)^2} \, dr
\le C + C/\gamma.
\end{align*}
Using wave operator for $-\Delta + V$, we have similar estimates
if $p^2$ is replaced by $-\Delta + V$. Since $\la \phi_0
Q_1^2=O(n^2)$,
\[
(h_*,h_*) \le C n^2\gamma^{-1} n^2 \le C,
\]
where $C$ is independent of $n$.
Since $u = \gs + \bar h= \phi_0 + \bar h + O(n^2)$, we have
obtained the $u$ part of the estimates $\norm{\evector}_{L^2} \le
C$ and \eqref{3-18}. The corresponding estimate for $v$ can be
proved using $v = (-L_+)u/\evalue$.

\subsection{Resolvent estimates}

In this subsection we study the resolvent $R(w) = (w - \L_1)
^{-1}$. Note that $R(w)$ had a different meaning in the previous
subsection. We will prove resolvent estimates along the continuous
spectrum $\Sigma_c$ and determine all eigenvalues outside of
$\Sigma_c$.


Let $L^2_r$ denote the weighted $L^2$ spaces for $r \in \R$:
\[
L^2_r = \bket{f: (1+x^2)^{r/2} f(x) \in L^2(\R^3)}.
\]
We will prove the
following lemma.

\begin{lemma} \label{th:resolvent}
Let $R(w)= (w - \L_1)^{-1}$ be the resolvent of $\L_1$. Let ${\bf
B}=B(L^2_r,L^2_{-r})$, the space of bounded operators from $L^2_r$
to $L^2_{-r}$ with $r>3$. Recall $\evalue = i \ka + \gamma$. For
$\tau \ge |E_1|$ we have
\begin{equation}
\norm{R(i \tau \pm 0)}_{\bf B}+ \norm{R(-i \tau \pm 0)}_{\bf B}
\le C (1+\tau)^{-1/2}+ C (|\tau - \ka| + n^4)^{-1}.
\end{equation}
The constant $C$ is independent of $n$. We
also have
\begin{equation} \norm{R^{(k)}(i \tau \pm 0)}_{\bf B}+
\norm{R^{(k)}(-i \tau \pm 0)}_{\bf B} \le C (1+\tau)^{-(1+k)/2}+ C
(|\tau - \ka| + n^4)^{-1}.
\end{equation}
for derivatives, where $k=1,2$.
\end{lemma}

We first consider $R_0(w)=(w-JH_1)^{-1}$. Recall $H_1 = - \Delta +
V -E_1$. Since
\begin{align}\nonumber
(w-JH_1)^{-1} &= \begin{bmatrix}w&-H_1\\H_1&w\end{bmatrix}^{-1}
=\frac 1{H_1^2 + w^2}\begin{bmatrix} w&H_1\\-H_1 &w\end{bmatrix}
\\
\label{JH-resol} &=\frac 12 \begin{bmatrix} -i&1\\-1
&-i\end{bmatrix}(H_1-i w)^{-1} + \frac 12 \begin{bmatrix} i&1\\-1
&i\end{bmatrix}(H_1+i w)^{-1},
\end{align}
the estimates of $R_0(w)$ can be derived from that of
$(H_1-iw)^{-1}$ and $(H_1+iw)^{-1}$. By assumption, the bottom of
the continuous spectrum of $H_1$,  $-E_1$, is not an eigenvalue
nor a resonance of $H_1$. Hence $(H_1-z)^{-1}$ is uniformly
bounded in ${\bf B}$ for $z$ away from $e_0-E_1$ and $e_1-E_1$,
see \cite{JK}. By \eqref{JH-resol} and \eqref{2-3}, $R_0(w)$ is
uniformly bounded in ${\bf B}$ for $w$ with $\dist(w,\Sigma_p)\ge
n$, where $\Sigma_p=\bket{0, i e_{01}, -i e_{01} }$.

Write
\[
\L_1 = JH_1 + W, \qquad W=\begin{bmatrix} 0& \la Q_1^2 \\ -3\la
Q_1^2 & 0 \end{bmatrix}.
\]
For $R(w)=(w - \L_1)^{-1}$ we have
\begin{equation} \label{3-40} R(w) = (1 - R_0(w)W)^{-1} R_0(w) =
\sum_{k=0}^\infty [R_0(w)W)]^k R_0(w). \end{equation}
Since $R_0(w)$ is uniformly bounded in ${\bf B} $ for $w$  with
$\dist(w,\Sigma_p)>n$, and $W$ is localized and small,
\eqref{3-40} converges and $(w - \L_1)^{-1}$ is uniformly bounded
in ${\bf B} $ for $w$ with $\dist(w,\Sigma_p)>n$ and we have
\begin{equation} \label{3-21N} \norm{R(w)}_B \le C \dist (w, \Sigma_p)^{-1},
\qquad (n\le \dist(w,\Sigma_p)\le 1). \end{equation}

Recall $\Sigma_c=\bket{is: |s|\ge |E_1|}$ is the continuous
spectrum of $JH_1$ and $\L_1$. For $w$ in the region
\begin{equation} \label{3-41} \bket{w: \dist (w,\Sigma_p)\ge n, w \not \in
\Sigma_c}, \end{equation}
we have
\[
\norm{R_0(w)}_{(L^2,L^2)} \le C\dist (w,\Sigma_c)^{-1}.
\]
By \eqref{3-40}, and because $W$ is localized and small,
\begin{align*}
&\norm{R(w)}_{(L^2,L^2)} \le \norm{R_0(w)}_{(L^2,L^2)} +
\\
&\qquad +\sum_{k=1}^\infty C\norm{R_0(w)}_{(L^2,L^2)}
\bket{C n^2 \norm{R_0(w)}_{\bf B}}^{k-1}\norm{R_0(w)}_{(L^2,L^2)} \\
& \le C\dist (w,\Sigma_c)^{-1}+ C\dist (w,\Sigma_c)^{-2}.
\end{align*}
Hence $R(w)$ is uniformly bounded in $(L^2,L^2)$ in a neighborhood
of $w$. In particular, there is no eigenvalue of $\L_1$ in the
above region \eqref{3-41}. Note that this region includes a
neighborhood of the bottom of the continuous spectrum $\Sigma_c$,
$\pm i E_1$, except those in $\Sigma_c$. Hence the eigenvalues can
occur only in $\bket{w: \dist (w,\Sigma_p)< n}$ or $\Sigma_c$.

The circle $\bket{w:|w|=\sqrt n}$ is in the resolvent set of
$\L_1$. By \cite{RS} Theorem XII.6, the Cauchy integral
\[
P= \frac 1{2 \pi i} \oint_{|w|=\sqrt n} (w -\L_1)^{-1} \, d w
\]
gives the $L^2$-projection onto the generalized eigenspaces with
eigenvalues inside the disk $\bket{w:|w|<\sqrt n}$. Moreover, the
dimension of $P$ is an upper bound for the sum of the dimensions
of those eigenspaces. However, since the projection $P_0=(2 \pi
i)^{-1} \oint_{|w|= \sqrt n}  R_0(w) \, dw$ has dimension $2$ (see
\eqref{2-2A}--\eqref{2-3}), and
\[
P -P_0 = \frac 1{2 \pi i} \oint_{|w|=\sqrt n} \sum_{k=1}^\infty
\,[R_0(w)W]^k \, R_0(w) \, d w
\]
is convergent and bounded in $(L^2,L^2)$ by
\begin{align*}
&\le C \norm{R_0(w)}_{(L^2,L^2)} n^2 \sum_{k=0}^\infty \bke{ C n^2
\norm{R_0(w)}_{\bf B} }^k \norm{R_0(w)}_{(L^2,L^2)}
\\
&\le C n^{-1/2} C n^2 n^{-1/2}= C n,
\end{align*}
(here we have used \eqref{3-21N}), the dimension of $P$ is also
two. Since we already have two generalized eigenvectors
$\svect{0\\Q_1}$ and $\svect{R_1\\0}$ with eigenvalue $0$, we have
obtained all generalized eigenvectors with eigenvalues in the disk
$|w|<\sqrt n$. Together with the results in \S 2.1, we have
obtained all eigenvalues outside of $\Sigma_c$: $0$, $\pm \evalue$
and $\pm \wbar \evalue$.

We next study $R(w) = (w - \L_1)^{-1}$ for $w$ near $\pm i
e_{01}$: $|w- i e_{01}|<n$ or $|w+ i e_{01}|<n$. Let us assume $w
= i \tau - \e$ with $\tau, \e>0$, thus $- w^2$ lies in $G$
(defined in \eqref{G.def}). The other cases are similar. Let $\svect{f \\
g} \in \Complex L^2$. We want to solve the equation
\begin{equation}
\label{2-48} (w - \L_1) \vect{ u \\ v} = \vect{ f \\ g}.
\end{equation}
We have
\[
w u -  H v = f, \qquad w v + (H + 2 \la Q_1^2) u = g.
\]
Cancelling $v$, we get (recall $A=H 2\la Q_1^2$)
\[
w^2 u + (H^2 + A)u = F, \qquad F = w f + H g.
\]

Write $u = \al \gs + \beta \es + \eta$ with $\eta \in \Hc(H)$
and $\es =Q_1/\norm{Q_1}_2$.
Also denote $\zeta =\al \gs + \beta \es$. We have
\begin{align*}
\bke{w^2  + H^2 + \Pc A} \eta &= \Pc F - \Pc A \zeta,
\\
\bke{w^2  + H^2 + {\bf P}^\perp A} \zeta &= {\bf P}^\perp F - {\bf P}^\perp A \eta.
\end{align*}
Here $\Pc =\Pc(H)$ and ${\bf P}^\perp = 1 - \Pc$. Solving $\eta$
in terms of $\zeta$, we get
\begin{equation} \label{5-19} \eta =\bke{w^2  + H^2 + \Pc A \Pc}^{-1}
(\Pc F - \Pc A \zeta). \end{equation}
Substituting the above into the $\zeta$ equation we get
\begin{equation} \label{5-20}
\bke{w^2  + H^2 + {\bf P}^\perp A - {\bf P}^\perp
A \bke{w^2  + H^2 + \Pc A \Pc }^{-1}\Pc A} \zeta = \wt F ,
\end{equation}
\[
\wt F= {\bf P}^\perp F - {\bf P}^\perp A \bke{w^2  + H^2 + \Pc A
\Pc}^{-1} \Pc F.
\]
Using $\gs$ and $\es$ as basis, we can put \eqref{5-20} into
matrix form
\begin{equation} \label{5-21}
\begin{bmatrix}
a & b \\
0 & w^2
\end{bmatrix}
\vect{\al \\ \beta} = \vect{(\gs, \wt F) \\ (\es, \wt F)},
\end{equation}
where (recall $H\gs = - \rho \gs$, $H \es = 0$)
\[
a=w^2 + \rho^2 - \rho(\gs 2 \la Q_1^2 \gs) + \rho (\gs 2 \la Q^2,
\bke{w^2  + H^2 + \Pc A \Pc }^{-1} H \Pc 2 \la Q_1^2 \gs),
\]
\[
b =- \rho(\gs 2 \la Q_1^2 \es) + \rho (\gs 2 \la Q^2, \bke{w^2  +
H^2 + \Pc A \Pc }^{-1} H \Pc 2 \la Q_1^2 \es).
\]
Thus
\begin{equation}
\vect{\al \\ \beta} = \begin{bmatrix}
1/a & -b/(aw^2) \\
0 & w^{-2}
\end{bmatrix}
\vect{(\gs, \wt F) \\ (\es, \wt F)}.
\end{equation}

\bigskip

We now consider the case when $w$ is near the continuous spectrum
$\Sigma_c$. We will assume $w= i\tau - \e$ with $|\tau -
e_{01}|<n$ and $\e>0$ much smaller. The case $w= i\tau + \e$
follows similarly.

Let $z = - w^2$. It follows that $z\in G$ and $\Re z>0$ is small.
The idea of what follows is to compare $z$ with $z_*$, the fixed
point found in \S 2.1.

We have $a=- z + f(z)= (z_* -z) + (-f(z_*) + f(z))$. Using Lemma
\ref{th:A-1} (3) with $w_1=z$ and $w_2=z_*$, we have
\[
|a|\ge |z-z_*| - |-f(z_*) + f(z)| \ge \tfrac 12 |z-z_*|= \tfrac 12
|w^2-\bar \evalue ^2| \ge C |w- \bar \evalue|.
\]
Recall $\evalue = i \ka + \gamma$ with $\gamma \sim n^4$. Hence
$|a|\ge  C (|\tau - \ka| + n^4)$. Thus
\[
|\al|+ |\beta| \le C(1 + |a|^{-1}) \| \wt F\|_{L^2} \le C(|\tau -
\ka| + n^4)^{-1} \bke{\norm{ f}_{L^2}+ \norm{g}_{L^2}}.
\]

By \eqref{5-19} and $F=wf + H g$,
\[
\eta = \Omega w \Pc f + \Omega H \Pc g - \Omega w \Pc A \zeta,
\]
where $\Omega =\bke{w^2  + H^2 + \Pc A \Pc}^{-1}$.
Substituting the above into \eqref{5-19}, we can solve $\eta$ and we have
\[
\norm{\eta}_{L^2} \le C \bke{\norm{ f}_{L^2}+ \norm{g}_{L^2}} + C
n^2 (|\tau - \ka| + n^4)^{-1}\bke{\norm{ f}_{L^2}+
\norm{g}_{L^2}}.
\]
We conclude, for $u = \al \gs + \beta \es + \eta$,
\[
\norm{u}_{L^2} \le \bke{ C + C (|\tau - \ka| + n^4)^{-1}}
\bke{\norm{ f}_{L^2}+ \norm{g}_{L^2}}.
\]
We can estimate $v$ similarly. Thus, for $\tau \in (e_{01}-n,
e_{01}+ n$),
\[ 
\norm{R(i \tau \pm 0)}_{\bf B} \le
 C + C (|\tau - \ka| + n^4)^{-1}.
\]

For $\tau > e_{01}+ n $ and $w=i\tau + 0$,  using $R(w)= (1 +
R_0(w ) W)^{-1} R_0(w)$ and the fact that $\norm{R_0(w)}_{\bf B}
\le C (1+\tau)^{-1/2}$, (see \cite{JK} Theorem 9.2), we have
$\norm{R(i\tau +  0)}_{\bf B} \le C \tau^{-1/2}$.

For $\tau \in [|E_1|, e_{01}-n]$, by the same argument we have
$\norm{R(i\tau + 0)}_{\bf B} \le C$.

The derivative estimates for the resolvent is obtained by
induction argument and by differentiating the relation $R(1+W
R_0)=R_0$ and using the relations $(1+W R_0)^{-1} = 1- WR$ and
$(1+ R_0 W)^{-1} = 1- RW$. See the proof of \cite{JK} Theorem 9.2.
We have proved Lemma \ref{th:resolvent}.

\subsection{Nonexistence of generalized $\evalue$-eigenvector}

We now show that $\evalue$ is simple and $\evector$ is the only
generalized $\evalue$-eigenvector, i.e., there is no vectors
$\phi$ with $(\L_1 - \evalue) \phi\not =0$ but $(\L_1 - \evalue)^k
\phi=0$ for some $k\ge 2$. Suppose the contrary, then we may find
a vector
$\vect{u\\v} $ with $(\evalue - \L_1) \vect{u\\v} =\vect{u_* \\
v_*}$. That is, $w=\evalue$ and $\vect{f\\g} = \vect{u_* \\ v_*}$
in the system \eqref{2-48}. We have $F= w u_* + H v_* = 2 \evalue
u_*$. Since $u_* = \gs + \bar h_*$ with $\bar h_* \in \Hc(H)$, we
have $(\es,\wt F)= (\es,F) = (\es,2\evalue u_*)=0$. Hence
$\beta=0$. Also
\begin{align*}
 (\gs,\wt F)&=(\gs,F) - (\gs H 2\la Q_1^2
\bke{w^2  + H^2 + \Pc A \Pc}^{-1} \Pc F)
\\
&= 2 \evalue + \rho (\gs 2 \la Q_1^2\bke{w^2  + H^2 + \Pc A
\Pc}^{-1} 2 \evalue \bar h_*)
\\
&= 2\evalue \bkt{ 1 +\rho (\Phi, \Omega \wbar \Omega H \Phi)},
\end{align*}
where $\Omega =\bke{w^2  + H^2 + \Pc A \Pc}^{-1}$ and $\Phi=\Pc
\gs 2 \la Q_1^2$. Since the main term in $(\Phi, \Omega \wbar
\Omega H \Phi)$,
\[
(\Phi, (w^2  + H^2 )^{-1} (\bar w^2  + H^2)^{-1} H \Phi),
\]
is positive, $(\gs,\wt F)$ is not zero.

On the other hand, $a= \evalue^2 + f(- \evalue^2) = -\bar z_* +
f(\bar z_*)=0$. Hence there is no solution for $\al$. This shows
$\evalue$ is simple (and so are $-\evalue, \pm \bar \evalue$).

Once we have an eigenvector $\evector$ with $\L_1 \evector =
\evalue \evector$ and $\evalue $ complex, then we have three other
eigenvalues and eigenvectors as given in \eqref{3-17}. Hence we
have found all eigenvalues and eigenvectors of $\L_1$. $\Complex
\eigen$ is the combined eigenspace of $\pm \evalue$ and $\pm \bar
\evalue$. It is easy to check that  $\RE \Complex \eigen =
\eigen$.  We have proved parts (1)--(3) of Theorem \ref{th:2-2} in
\S 2.1 to \S 2.3.

\subsection{Nonexistence of embedded eigenvalues}
\label{sec:embedded}

In this subsection we prove that there is no embedded eigenvalue
$i \tau$ with $|\tau|>|E_1|$. Suppose the contrary, we may assume
$\tau > - E_1>0$ and $\L_1 \psi = i \tau \psi$ for some $\psi \in
\Complex L^2$. We will derive a contradiction.

Let $H_* = - \Delta - E_1$. We can decompose
\begin{equation} \label{L-Hstar} \L_1 = JH_* + A, \qquad A= \begin{bmatrix} 0 &
V + \la Q_1^2 \\ - V - 3 \la Q_1^2 & 0
\end{bmatrix}.
\end{equation}
Hence $(i \tau - JH_*)\psi = A \psi$. By the same computation of
\eqref{JH-resol} we have
\[
(w -JH_*)^{-1}=(H_*-i w)^{-1}M_+ + (H_*+ i w)^{-1} M_-,\quad
\]
where
\[
M_+=\frac 12 \begin{bmatrix} -i&1\\-1 &-i\end{bmatrix} ,
\qquad M_-= \frac 12 \begin{bmatrix} i&1\\-1 &i\end{bmatrix}.
\]
Thus, with $w = i\tau$, we have
\begin{equation} \label{2-55} \psi =(i \tau - JH_*)^{-1} A \psi =
(H_*+\tau)^{-1}\phi_+ + (H_*-\tau)^{-1} \phi_-  ,
\end{equation}
where $\phi_+ =M_+A \psi $ and  $\phi_- =M_- A \psi $.  By
Assumption A1 on the decay of $V$ and that $\psi \in L^2$,  both
$\phi_+,\phi_- \in L^2_{5+\sigma}$ with $\sigma>0$. Since $- \tau$
is outside the spectrum of $H_*$, we have $(H_*+\tau)^{-1}\phi_+
\in L^2_{5+\sigma}$. Let $s=E_1 + \tau>0$. We have $H_*-\tau =
-\Delta - s$. By assumption $\psi \in \Complex L^2$, hence so is
$(H_*-\tau)^{-1} \phi_- $. Therefore $(p^2 - s)^{-1}
\widehat{\phi_-}(p)\in L^2$. Since $\phi_- \in L^2_{5+\sigma}$,
$\phi_-$ is continuous and we can conclude
\begin{equation} \label{Restrict0}
\widehat{\phi_-}(p)\big|_{|p|=\sqrt s} = 0.
\end{equation}

We now recall \cite{RS2} page 82, Theorem IX.41: Suppose $f \in
L^2 _r$ with $r>1/2$ and let $B_s f =\bke{(p^2 - s)^{-1}
\widehat{f}} ^\vee $. Suppose $\hat f (p)\big|_{|p|=\sqrt s } =
0$. Then for any $\e>0$, one has $B_s f \in L^2_{r - 1 -2 \e}$ and
$\norm{B_s f }_{ L^2_{r - 1 - 2 \e}}\le C_{r,\e,s}
\norm{f}_{L^2_r}$ for some constant $C_{r,\e,s}$.

In our case, we have $f=\phi_-$, $\e= \sigma/2$ and $r=5+\sigma$.
We conclude $(H_*-\tau)^{-1} \phi_- =B_s f \in L^2_{4}$. Thus
$\psi \in L^2_{4}$.

However, since $(z-\L_1)\psi = (z-i\tau)\psi$, we have $R(z) \psi
= (z-i\tau)^{-1} \psi$. Thus we have
\[
\norm{(z-i\tau)^{-1}  \psi}_{L^2_{-r}} \le C \norm{
\psi}_{L^2_{4}},
\]
where the constant $C$ remains bounded as $z \to i \tau$ by Lemma
\ref{th:resolvent}. This is clearly a contradiction. Thus $\psi$
does not exist.

\subsection{Absence of eigenvector and resonance at bottom of
continuous spectrum}

In this subsection we show that there is no eigenvector and
resonance at $\pm i E_1$. We want to show that, for
$n=\norm{Q_{1,E_1}}_{L^2}$ sufficiently small, the null space of
$\L_1 \pm i E_1$ in $X = L^2_{-r}$, $ r> 1/2$, is zero. Let us
consider the case at $i|E_1|$. Suppose otherwise, we have a
sequence $Q_{1,E_1(k)} \to 0$ and $\psi _k$ so that
\[
\bke{\L_{1,E_1(k)} + i E_1(k)}\, \psi_k = 0, \quad \norm{\psi_k}_X
=1.
\]
As in the previous subsection, we write $\L_{1,E_1(k)} = JH_* +
A_k$, where $H_*= -\Delta -E_1(k)$ and $A_k = JV + \begin{bmatrix}
0 & 1 \\ -3 & 0\end{bmatrix} \la Q_{1,E_1(k)}^2$. By \eqref{2-55}
with $\tau = |E_1(k)|$ we have
\[
\psi_k =(i \tau - JH_*)^{-1} A_k \psi_k = (- \Delta +
2\tau)^{-1}M_+A_k \psi_k + (- \Delta)^{-1} M_- A_k   \psi_k
\]
in $X$. Note that $(- \Delta + 2\tau)^{-1}M_+A_k$ and $(- \Delta)
^{-1} M_- A_k$ are compact operators in $X$, with a bound uniform
in $k$. Since $X$ is a reflexive Banach space, we can find a
subsequence, which we still denote by $\psi_k$, converging weakly
to some $\psi_* \in X$. Thus $\tau \to |e_1|$, $(- \Delta +
2\tau)^{-1}M_+A_k \psi_k \to (- \Delta  - 2e_1)^{-1}M_+JV \psi_*$
and $ (- \Delta)^{-1} M_- A_k \psi_k \to (- \Delta)^{-1}M_+JV
\psi_*$ strongly in X. Thus
\[
\psi_* =(- \Delta - 2e_1)^{-1}M_+JV \psi_* + (- \Delta)^{-1}M_+JV
\psi_*
\]
and $\psi_k \to \psi_*$ strongly. Hence $\norm{\psi_*}_X = \lim
\norm{\psi_k}_X = 1$ and $(JH_1 + i e_1)\psi_*=0$ by \eqref{2-55}
again. This contradiction to our assumption shows our claim.

\bigskip

{\bf Another proof:}

We will use the resolvent estimates Lemma \ref{th:resolvent} to
give a proof, without using that $\L_1$ is a perturbation of
$JH_1$. Suppose the contrary that we have $\psi \in X$ which
satisfies $\psi \not = 0$, $\L_1 \psi = i \tau \psi$, with $\tau =
|E_1|$. Write $\L_1 = JH_*+A$ as before and let $R(z) = (z -
\L_1)^{-1}$ and  $R_0(z) = (z - JH_*)^{-1}$. We have $(i \tau -
JH_*)\psi = A\psi$, hence
\begin{equation} \label{2-R0A} \psi =
R_0(i \tau) A\psi \qquad \text{in } X.
\end{equation}

Let $w= \sigma_1 A \psi = A^*  \sigma_1 \psi$. (Note $A^* =
\sigma_1 A \sigma_1$). We have that $Vw \in L^2_{r}$. By Lemma
\ref{th:resolvent} the ${L^2_{-r}}$-norm of $R(z) Vw $ is
uniformly bounded as $z \to i \tau$. We will derive a
contradiction.

Recall the resolvent identity  $R(z) A = - 1 - (1 -
R_0(z)A)^{-1}$. Hence $ \rho(z) \equiv (1 - R_0(z)A)^{-1} w =
R(z) A  w + w$ is also uniformly bounded in $L^2_{-r}$ as $z \to i
\tau$.

Recall \cite{JK} Lemma 2.3 that $(- \Delta-z)^{-1} =
(-\Delta)^{-1} + O(\sqrt{|z|})$ in ${\cal B} \bke{H^{-1}_{s},
H^1_{-s}}$ for $s>1$. Hence for $z$ near $i \tau$ we have by
\eqref{2-55}
\[
R_0(z) = R_0( i \tau) + O(\sqrt{|z-i\tau|})
\]
in ${\cal B}\bke{H^{-1}_{s},H^1_{-s}}$. Therefore
\begin{align*}
(w,w)&=(A^*\sigma_1 \psi,\ (1-R_0(z)A) \rho(z))
\\
&=(A^*\sigma_1 \psi, \ [1 - R_0(i \tau) A] \rho(z)) + (A^*\sigma_1
\psi, O(\sqrt{|z-i\tau|})A\rho(z)).
\end{align*}
Since $\sigma_1 J = - J \sigma_1$, $\sigma_1 (z-JH_*) =  (z +JH_*)
\sigma_1 =  (z -JH_*)^* \sigma_1$. Hence $\sigma_1 R_0(i \tau) =
R_0(i \tau)^* \sigma_1$ and we have by \eqref{2-R0A}
\[
\sigma_1 \psi = \sigma_1 R_0(i \tau) A\psi= R_0(i \tau)^* \sigma_1
A\psi= R_0(i \tau)^*  A^*\sigma_1\psi.
\]
Hence
\[
(A^*\sigma_1 \psi, \  R_0(i \tau) A \rho(z)) = ( A^* R_0(i \tau)^*
A^*\sigma_1 \psi, \ \rho(z)) = ( A^* \sigma_1 \psi, \ \rho(z)).
\]
Hence $(A^*\sigma_1 \psi, \ [1 - R_0(i \tau) A] \rho(z))=0$. Since
$\rho(z)$ is uniformly bounded in $L^2_r$, we have
\[
(w,w) =(A^*\sigma_1 \psi, O(\sqrt{|z-i\tau|})A\rho(z))=
O(\sqrt{|z-i\tau|})
\]
as $ z \to  i \tau$. Thus $w=0$. Hence $(i \tau - JH_*)\psi =0$.
If we write $\psi = \svect{u\\v}$, then
\[
i \tau u - (-\Delta + \tau)v =0, \quad (-\Delta + \tau) u +i \tau
v =0.
\]
One gets $\Delta (u -i v)=0$ immediately. Since $u,v\in X$, we
conclude $u=iv$ and $(-\Delta + 2\tau)u =0$. Hence $u,v=0$. This
finishes the proof.

\subsection{Proof of Theorem \ref{th:2-2} (4)--(6)}

We first show the orthogonality conditions. Recall
$\sigma_1 = \begin{bmatrix}0 & 1 \\ 1 & 0\end{bmatrix}$. It is
self-adjoint in $\Complex L^2$. Let
$\L_1^*$ be the adjoint of $\L_1$ in $\Complex L^2$. We have $\L_1
^*= \begin{bmatrix}0 & -L_+ \\ L_- & 0\end{bmatrix}$ and $\L_1
^*=\sigma_1 \L_1 \sigma_1$.

Suppose $\L_1 f = \om _1 f$ and $\L_1 g = \om _2 g$ with $ \bar\om
_1 \not = \om _2$. We have $\L_1^* \sigma_1 f =\sigma_1 \L_1  f=
\om _1 \sigma_1 f$ and hence
\[
\om _2 (\sigma_1 f, g) = (\sigma_1 f, \om _2 g) = (\sigma_1 f,
\L_1 g) = (\L_1^*\sigma_1 f, g) = (\om _1 \sigma_1 f, g)= \bar \om
_1( \sigma_1 f, g).
\]
Hence we must have $(\sigma_1 f, g) =0$. Therefore we have
$\sigma_1 \bar \evector \perp \bar \evector, \sigma_3 \evector,
\sigma_3 \bar \evector$, $\sigma_1  \evector \perp  \evector,
\sigma_3 \evector, \sigma_3 \bar \evector$, etc. If we write
$u=u_1 + i u_2$, $v=v_1+i v_2$ and $\evector= \svect{u
\\v}$, then we have
\begin{equation} \label{3-24} \int \bar u v \,d x = 0.
\end{equation}
In other words, $(u_1,v_1)+(u_2,v_2)=0$ and $(u_1,v_2)= (u_2,v_1)$.

If $f \in S(\L_1)$ and $\L_1 g = \om _2 g$ with $\om _2 \not = 0$.
We have $(\L_1^*)^2\sigma_1 f = 0 $, hence
\[
(\sigma_1 f, \om _2^2 g) = (\sigma_1 f, \L_1^2 g) =
((\L_1^*)^2\sigma_1 f, g) = (0, g).
\]
Hence $(\sigma_1 f,  g)=0$. In terms of components, we get
$(Q_1,u_1)=(Q_1,u_2)=0$, $(R_1,v_1)=(R_1,v_2)=0$. The above shows
\eqref{3-21}. The rest of (4) and (5) follows directly.

To prove (6), we first prove the following spectral gap
\begin{equation} \label{gap2}
L_+|_{\bket{Q_1,v_1,v_2}^\perp} > \tfrac 12 |e_1| ~, \qquad
L_-|_{\bket{R_1,u_1,u_2}^\perp} > \tfrac 12 |e_1| ~.
\end{equation}
We will show the first assertion. Note that by \eqref{3-18} we
have
\[
v_1 = \Pc(L_-) v_1 + O(n^2), \qquad v_2 = - \phi_0 + \Pc(H_1) v_2
+ O(n^2)
\]
in $L^2$. In particular $\norm{v_2}_{L^2} \ge 1/2$, and $ (v_1,L_-
v_1)\ge ( v_1, L_- \Pc (L_-)v_1)  - C n^2\ge - Cn^2$. By
\eqref{3-24}
\[
(v_1,L_- v_1) + (v_2,L_- v_2) =(v, L_-v )= (v,\om u) =0.
\]
Hence $(v_2,L_+v_2) = (v_2, L_-v_2 ) + O(n^2) \le C n^2 $. We also
have $(Q_1,L_+ Q_1)=(Q_1,L_- Q_1) + O(n^4) =0+ O(n^4)$. Let $Q_1'=
Q_1 - (Q_1,v_2)v_2 /\norm{v_2}_2^2$. We have $Q_1' \perp v_j$ and
$Q_1'= Q_1 + O(n^3)$ by \eqref{3-18} again. Hence $(Q_1',L_+
Q_1')=(Q_1,L_+ Q_1)+O(n^4)=O(n^4)\le C n^2 (Q_1',Q_1')$. We
conclude that $L_+ |_{\myspan \bket{Q_1,v_2}} \le C n^2$. Since
$L_+$ is a perturbation of $H_1$, it has exactly two eigenvalues
below $\tfrac 12 |e_1|$. By minimax principle we have
$L_+|_{\bket{Q_1,v_2}^\perp} > \tfrac 12 |e_1|$. This shows the
first assertion of \eqref{gap2}. The second assertion is proved
similarly.

Let $\bQ(\psi)$ denote the quadratic form: (see e.g. \cite{W1,W2})
\begin{equation} \label{bQ.def} \bQ(\psi)=( f, L_+ f) + ( g, L_- g)~, \qquad
\text{if } \psi=f + i g ~.
\end{equation}
One can show for any $\psi \in L^2$
\begin{equation}  \label{3-13} \bQ(e^{t\L_1}\psi)=\mathbf{Q}(\psi)~, \qquad
\text{for all } t, \end{equation}
by direct differentiation in $t$. By \eqref{gap2} one has
\[
\bQ(\eta) \sim \norm{\eta}_{H^1}^2 ~, \qquad
\text{for any } \eta \in \Hc(\L_1 ) .
\]
Thus
\[
\norm{e^{t\L_1}\eta}_{H^1}^2 \sim \bQ(e^{t\L_1}\eta)=
\bQ(\eta)\sim \norm{\eta}_{H^1}^2.
\]
Similarly, we have by \eqref{gap2} and the above relation
\[
\norm{\eta}_{H^3}^2 \sim \norm{\L_1 \eta}_{H^1}^2 \sim \bQ(\L_1 \eta).
\]
Since $\bQ(\L_1 \eta)=\bQ(e^{t\L_1}\L_1 \eta)$, we have
$\norm{\eta}_{H^3} \sim \norm{e^{t\L_1} \eta}_{H^3}$. By interpolation
we have $\norm{\eta}_{H^2} \sim \norm{e^{t\L_1} \eta}_{H^2}$.
We have proven (6).

\subsection{Wave operator and decay estimate}

It remains to prove the decay estimate (7). We will use the wave
operator. We will compare $\L_1$ with $J H_*$, where $H_* =-\Delta
- E_1$. Recall we write $\L_1 = JH_* + A$ in \S
\ref{sec:embedded}, \eqref{L-Hstar}.
Keep in mind that $H_*$ has no bound states and $A$ is local.
Define $W_+ = \lim_{t \to +\infty} e^{-t\L_1}e^{tJH_*}$. Let
$R(z)=(z - \L_1)^{-1}$ and $R_*(z) = (z -JH_*)^{-1}$. We have
\begin{align*}
&W_+ f -f
\\
&= \lim_{\e \to 0+} \int_{|E_1|}^{+\infty}  R(i\tau + \e) A
\bkt{R_*(i\tau -\e)- R_*(i \tau + \e) } f \, d \tau
\\
& - \lim_{\e \to 0+} \int_{|E_1|}^{+\infty}  R(-i\tau + \e) A
\bkt{R_*(-i\tau -\e)- R_*(-i \tau + \e) } f \, d \tau.
\end{align*}

Yajima \cite{Y,Y2} was the first to give a general method for
proving the $(W^{k,p},W^{k,p})$ estimates for the wave operators
of self-adjoint operators. This method was extended by Cuccagna
\cite{C} to non-selfadjoint operators in the form we are
considering. (He also used idea from Kato \cite{K}). One key
ingredient in this approach is the resolvent estimates near the
continuous spectrum, which in many cases can be obtained by the
Jensen-Kato \cite{JK} method. (See \cite{Y} Lemmas 3.1 and 3.2 and
\cite{C} Lemmas 3.9 and 3.10).
In our current setting, this estimate is provided by the Lemma
\ref{th:resolvent}. We can thus follow the proof of \cite{C} to
obtain that $W_+ $ is an operator from $\Complex L^2$ onto
$\Hc(\L_1)$. Furthermore,  $W_+$ and its inverse (restricted to
$\Hc(\L_1)$) are bounded in $(L^p,L^p)$-norm for any $p\in
[1,\infty]$. (Note this bound depends on $n$ since our bound on
$R(w)$ depends on $n$.) By the intertwining property of the wave
operator we have
\[
e^{t\L_1} \Pc = W_+ e^{t JH_*} (W_+)^*\Pc.
\]
 The decay estimate in (7) follows from the decay estimate of
$e^{t JH_*}$.

The proof of Theorem \ref{th:2-2} is complete.

\donothing{ Since \eqref{3-30} and \eqref{3-31} are similar, we
will only estimate \eqref{3-30}. Following \cite{Y}, we decompose
\eqref{3-30} as $W_1f + W_2f$, where
\begin{align*}
W_1 f&=\lim_{\e \to 0+} \int_{|E_1|}^{+\infty}  R_*(i\tau + \e) A
\bkt{R_*(i\tau -\e)- R_*(i \tau + \e) } f \, d \tau,
\\
W_2 f& =\lim_{\e \to 0+} \int_{|E_1|}^{+\infty} R_*(i\tau + \e) A
R(i\tau + \e) A \bkt{R_*(i\tau -\e)- R_*(i \tau + \e) } f \, d
\tau.
\end{align*}

The first term has an explicit form and can be estimated in $B(L^p,L^p)$
as in \cite{Y}.

The second term can be estimated directly using our previous estimates of
$R(w)$:
\begin{align*}
\norm{W_2}_{(L^p,L^p)} &\le \int _{|E_1|}^{+\infty}
\norm{ R_*(i\tau + \e)}_{(L^p,L^p)}
\norm{ R(i\tau + \e)}_{(L^2,L^2)}
\\
&\qquad \qquad \qquad \cdot
\norm{ R_*(i\tau -\e)- R_*(i \tau + \e) }_{(L^p,L^p)} \, d \tau
\\
&\le
\end{align*}
}

\subsection{Proof of Theorem \ref{th:2-1}}
\label{sec:pf21}

By the same Cauchy integral argument as in Subsection 2.2, the
only eigenvalues of $\L_1$ are inside the disks $\bket{w:|w|<
\sqrt n}$, $\bket{w:|w-ie_{01}|< \sqrt n}$ and
$\bket{w:|w+ie_{01}|< \sqrt n}$. Moreover, their dimensions are
$2$, $1$ and $1$, respectively, the same as that of $JH_1$. It
counts the dimension of (generalized) eigenspaces of $\L_1$ in
$\Complex L^2$. It also counts the dimensions of the restriction
of these spaces in $L^2=L^2(\R^3,\R^2)$ as a real-valued vector
space.

By \eqref{S.def}, we already have two generalized eigenvectors
near $0$. Hence we have everything near $0$.

Since the dimension is 1 near $i e_{01}$, there is only a simple
eigenvalue $\evalue$  near $i e_{01}$. We have $\evalue = i
e_{01}+O(n^2)$ since the difference between $\L_1$ and $JH_1$ is
of order $O(n^2)$. $\evalue$ has to be purely imaginary, otherwise
$- \bar \evalue$ is another eigenvalue near $ie_{01}$,
cf.~\eqref{3-17}, and the dimension can not be 1. (This also
follows from the Theorem of Grillakis.)

By the same arguments in \S 2.2-2.4 we can prove resolvent estimates
and the non-existence of embedded eigenvalues. Also, the bottoms of
the continuous spectrum are not eigenvalue nor resonance.

Let $\evector$ be an eigenvector corresponding to $\evalue $.
Since $\L_1 \evector = \evalue  \evector$ and $\bar \evalue  =
-\evalue $, we have $\L_1 \bar \evector = -\evalue  \bar
\evector$. Hence the (unique) eigenvalue near $-ie_{01}$ is
$-\evalue $
with eigenvector $\bar \evector$. Write $\evector=\svect{u\\
-iv}$. We may assume $u$ is real. Writing out $\L_1 \evector = i
\ka \evector$ we get $L_- v = -\ka u$ and $L_+ u = -\ka v$. Hence
$v$ is also real. We can normalize $u$ so that $(u,v)= 1$ or $-1$.
Since $\evector$ is a perturbation of $\svect{\phi_0\\ -i\phi_0}$,
we have $(u,v)= 1$.

With this choice of $u,v$, let $\Complex \eigen$ and $\eigen$ be
defined as in \eqref{E1.def}. $\Complex \eigen$ is the combined
eigenspace corresponding to $\pm \evalue $. Clearly $\RE \Complex
\eigen \subset \eigen$. Since
\[
a\vect{u\\0}+b\vect{0\\v} = \RE \al \evector , \qquad \al=a +b i,
\]
we have $\RE \Complex \eigen =\eigen$. That the choice of $\al$ is
unique can be checked directly. The statement that if $\zeta =\RE
\al \evector $ then  $\L_1 \zeta = \RE \evalue \al \evector $ and
$e^{t\L_1} \zeta = \RE e^{t\evalue } \al \evector $ is clear. We
have proved (3) and (4).

Clearly, $S(\L_1)$, $\eigen(\L_1)$ and $\Hc(\L_1)$ defined as in
\eqref{S.def}, \eqref{E1.def} and \eqref{Hc.def} are invariant
subspaces of $L^2$ under $\L_1$, and we have the decomposition
\eqref{L2.dec}. This is (2).

For (5), note that \eqref{Hc.def} is by definition. For
\eqref{SEorth}, we have
\begin{gather*}
(Q_1,u) = (Q_1, (-\ka)^{-1}L_- v) = (L_- Q_1, (-\ka)^{-1} v)=0,\\
(R_1,v) = (R_1, (-\ka)^{-1}L_+ u) = (-\ka)^{-1}(L_+ R_1,  u)=
(-\ka)^{-1} (Q_1, u) =0.
\end{gather*}
\eqref{3-10} comes from the orthogonal relations directly.

The first statement of (6) is because of (5). For the rest of (6),
We first prove the following spectral gap
\begin{equation} \label{gap1} L_+|_{\bket{Q_1,v}^\perp} > \tfrac 12 |e_1| ~,
\qquad L_-|_{\bket{R_1,u}^\perp} > \tfrac 12 |e_1| ~. \end{equation}
Since $L_+$ is a perturbation of $H_1$, it has exactly two
eigenvalues below $\tfrac 12 |e_1|$. Notice that $(Q_1, L_+Q_1) =
(Q_1L_-Q_1)+O(n^4) = O(n^4)$ and $(v,L_+v) = (v,-\ka u) = - \ka$.
Since $Q_1=n\phi_1 + O(n^3)$ and $v = \phi_0 + O(n^2)$, one has
$(Q_1,v)=O(n^3)$. Thus one can show $L_+|_{\myspan \bket{Q_1,v}}
\le C n^2$. If there is a $\phi \perp Q_1,v$ with $(\phi,L_+\phi)
\le \tfrac 12 |e_1|(\phi,\phi)$, then we have $L_+|_{\myspan
\bket{Q_1,v,\phi}} \le \tfrac 12 |e_1|$, which contradicts with
the fact that $L_+$ has exactly two eigenvalues below $\tfrac 12
|e_1|$ by minimax principle. This shows the first part of
\eqref{gap1}. The second part is proved similarly.

Recall the quadratic form $\bQ(\psi)$ defined in \eqref{bQ.def} in
\S 2.6. Also recall \eqref{3-13} that $\bQ(e^{t\L_1}\psi)=
\mathbf{Q}(\psi)$ for all $t$ and all $\psi \in L^2$. By the
spectral gap \eqref{gap1} one has
\begin{equation} \label{3-14} \bQ(\eta) \sim \norm{\eta}_{H^1}^2 , \quad 
\bQ(\L_1 \eta) \sim \norm{\eta}_{H^3}^2 , \qquad
\text{for any } \eta \in \Hc(\L_1 ) ~. \end{equation}

For $\psi \in M_1$, we can write $\psi=\zeta +\eta $, where $\zeta
=\RE \al \evector$, $\al\in \Complex$ and $\eta\in \Hc(\L_1 )$.
Notice that, by orthogonality in \eqref{Hc.def},
\[
\bQ(\psi) = - |\al|^2 \ka(u,v) + \bQ(\eta) ~,
\]
which is not positive definite, (recall $(u,v)=1$). However,
\begin{equation} \label{3-15} \norm{\psi}_{H^1}^2  \sim  |\al|^2 +
\norm{\eta}_{H^1}^2~. \end{equation}
To see it, one first notes that $\norm{\psi}_{H^1}^2 $ is clearly
bounded by the right side. Because of \eqref{3-10}, one has
$|\al|^2\le C\norm{\psi}_{H^1}^2$. One also has
$\norm{\eta}_{H^1}^2 \le C \norm{\phi}_{H^1}^2 + C|\al|^2$. Hence
\eqref{3-15} is true.

Therefore for $\psi=(\RE \al \evector) +\eta$ we have
\begin{alignat*}{2}
\norm{e^{t\L_1}\psi}_{H^1}^2  &\sim  \norm{e^{t\L_1}\RE \al
\evector}_{H^1}^2 + \norm{e^{t\L_1}\eta}_{H^1}^2 \qquad
&&\text{(by } \eqref{3-15})
\\
&\sim |e^{-it\evalue }\al |^2 + \bQ(e^{t\L_1}\eta) &&\text{(by
(4)}, \eqref{3-14})
\\
&\sim   |\al|^2 + \bQ( \eta) &&\text{(by \eqref{3-13})} ~.
\end{alignat*}
Hence we have $\norm{e^{t\L_1}\psi}_{H^1}^2 \sim
\norm{\psi}_{H^1}^2$ for all $t$. 
By an argument similar to that in \S 2.6, we have 
 $\norm{e^{t\L_1}\psi}_{H^k} \sim
\norm{\psi}_{H^k}$ for $k=3,2$. We have shown (6).

The decay estimate in (7) is obtained as in Theorem \ref{th:2-2}
(7). The constant $C$, however, is independent of $n$ in the
non-resonant case. The proof of Theorem \ref{th:2-1} is complete.

\section{Solutions converging to excited states}

In this section we prove Theorem \ref{th:1-2} using Theorems
\ref{th:2-1} and \ref{th:2-2}.  Since the proof for the non-resonant case
is easier, we will first prove the resonant case and then sketch the
non-resonant case. Note that we could follow the approach of
Theorem 1.5 of \cite{TY} if we had the transform $\L_1 \PcL = -U^{-1}
i A U \PcL$ as in \cite{TY}. However, it is not easy to define $A$ and $U$
for $\L_1$ and hence we choose another approach. This approach also gives
another proof for Theorem 1.5 of \cite{TY}.

Fix $E_1$ and $Q_1= Q_{1,E_1}$. Let $\L_1$ be the corresponding
linearized operator, and $\PM$, ${\bf P}_{\eigen}$ and
$\Pc^{\L_1}$ the corresponding projections with respect to
$\L_{1}$. For any $\xi_\infty \in \Hc(\L_1)$ with small $H^2 \cap
W^{2,1}$ norm, we want to construct a solution $\psi(t)$ of the
nonlinear Schr\"odinger equation \eqref{Sch} with the form
\[
\psi(t) = \bkt{ Q_1 + a(t)R_1+ h(t)  }e^{-i E_1 t+ i \theta(t)}~,
\]
where $a(t),\theta(t)\in \R$ and $h(t) \in M_1 = \eigen \oplus
\Hc(\L_1)$. Substituting the above ansatz into \eqref{Sch} and
using $\L_1 iQ_1=0$ and $\L_1 R_1= -i Q_1$, we get
\[
\pd_t h = \L_1 h + i^{-1} F(aR_1 +h) -i \dot \theta (Q_1 + aR_1 +
h) - a i Q_1 - \dot a R_1,
\]
where
\begin{equation}\label{F.def}
F(k)=\la Q_1(2|k|^2 + k^2) + \la |k|^2 k, \qquad k=aR_1+h.
\end{equation}
The condition $h(t)\in M_1$ can be satisfied by requiring that $h(0)\in
M_1$ and
\begin{align}
\label{a.eq} \dot a &=   (c_1 Q_1, \, \Im (F + \dot \theta h) ),
\\
\label{theta.eq} \dot \theta &= - [a + (c_1 R_1, \, \Re F) ]\cdot
[1+(c_1 R_1, R_1) a +(c_1 R_1,\Re h)]^{-1} ,
\end{align}
where $c_1= (Q_1,R_1)^{-1}$ and $F=F(aR_1+h)$. The equation for
$h$ becomes
\[
\pd_t h = \L_1 h + P_M F_{\text{all}}, \qquad F_{\text{all}} =
i^{-1} ( F+ \dot \theta (a R_1 +h)).
\]

The proofs of the two cases diverge here. For the resonant case we
decompose, using the decomposition of $M_1$ and \eqref{zeta.dec2}
of Theorem \ref{th:2-2},
\[
h(t) = \zeta(t)+ \eta(t),\qquad \zeta(t)= \RE \bket{ \al(t)
\evector + \beta(t) \sigma_3 \evector},
\]
where $\al(t),\beta(t) \in \Complex$ and $\eta(t) \in \Hc(\L_1)$.
Note
\[
\L_1 \zeta = {\bf RE} \bket{ \evalue \al \evector -\evalue  \beta
\sigma_3 \evector}.
\]
Recall  $\evalue = i \ka + \gamma$ with $\ka, \gamma >0$.
Taking the projections $P_\al$ and $P_\beta$  defined in
\eqref{PalPbeta.def} of Theorem \ref{th:2-2} of the $h$-equation,
we have
\begin{align}
\label{al.eq} \dot \al &= \evalue \al + P_\al F_{\text{all}},
\\
\label{beta.eq} \dot \beta &= -\evalue  \beta+ P_\beta
F_{\text{all}}.
\end{align}
Taking projection $\PcL$ we get the equation for $\eta$,
\[
\pd_t \eta = \L_1 \eta + \PcL i^{-1} \dot \theta \eta + \PcL \wt F, \qquad
\wt F =i^{-1} ( F+ \dot
\theta (a R_1 +\zeta)).
\]
We single out $\PcL i^{-1} \dot \theta \eta$ since it is a global linear term
in $\eta$ and cannot be treated as error. Let
\[
\wt \eta = \PcL e^{i \theta}  \eta .
\]
Note $\eta = \wt \eta + \PcL (1-e^{i \theta})\eta$ and $\Pc
(1-e^{i \theta})$ is a bounded map from $\Hc(\L_1)\cap H^2$ into itself
with its norm bounded by $C|\theta|$. Hence if $\theta$ is sufficiently
small, we can solve $\eta$ in terms of $\wt \eta$ by expansion:
\begin{equation}\label{solveeta}
\eta= U_\theta  \wt \eta, \qquad
U_\theta \equiv \sum_{j=0}^\infty \ [\Pc (1-e^{i \theta})]^j.
\end{equation}
The equation for $\wt \eta$ is
\begin{align*}
\pd_t \wt \eta &= \PcL e^{i \theta} (i \dot \theta \eta + \pd_t \eta)
\\
&=\L_1 \wt \eta + \bket{\PcL e^{i \theta} \L_1 - \L_1 \PcL e^{i \theta} }
\eta \\
&\quad+  \PcL e^{i \theta} \bket{
i \dot \theta \eta - \dot \theta \PcL i \eta + \PcL \wt F}
\end{align*}
Note that $i \dot \theta \eta - \dot \theta \PcL i \eta
= (1-\PcL)i \dot \theta \eta $ and
\begin{align*}
\bket{\PcL e^{i \theta} \L_1 - \L_1 \PcL e^{i \theta} }\eta
&= \PcL [e^{i \theta},\L_1]\eta
=\PcL  \sin \theta[i ,\L_1]\eta
\\& = \PcL  \sin \theta 2 \la Q_1^2 \bar \eta.
\end{align*}
Hence we have
\[
\pd_t \wt \eta = \L_1 \wt \eta + \PcL
\bket{\sin \theta 2 \la Q_1^2 \bar \eta
+ e^{i \theta}  (1-\PcL)i \dot \theta \eta +
e^{i \theta} \PcL \wt F }
\]

For a given profile $\xi_\infty$, let
\begin{equation}    \label{eq:9-4}
\wt  \eta(t)= e^{ t\L_1} \xii + g(t) .
\end{equation}
We have the equation
\begin{equation}    \label{g.eq}
\pd _t g  = \L_1  g + \PcL
\bket{\sin \theta 2 \la Q_1^2 \bar \eta
+ e^{i \theta}  (1-\PcL)i \dot \theta \eta +
e^{i \theta} \PcL \wt F } .
\end{equation}
We want $g(t)\to 0$ as $t \to \infty$ in some sense.

Summarizing, we write the solution $\psi(t)$ in the form
\begin{equation}  \label{psi.dec}
\begin{split}
\psi(t) &=  \Big \{ Q_1 + a(t)R_1+ \RE \bket{ \al(t) \evector +
\beta(t) \sigma_3 \evector} \\ & \qquad +  U_{\theta(t)}
(e^{ t\L_1} \xii + g(t)) \Big \} \, e^{-i E_1 t+ i \theta(t)}~,
\end{split}
\end{equation}
with $a(t)$, $\theta(t)$, $\al(t)$, $\beta(t)$ and $g(t)$
satisfying \eqref{a.eq}, \eqref{theta.eq}, \eqref{al.eq},
\eqref{beta.eq}, and \eqref{g.eq}, respectively.

The main term of $F$ is
\[
F_0 = \la Q_1 \bke{2|\xi|^2 + \xi^2}+\la |\xi|^2\xi, \qquad \xi(t)
= U_{\theta(t)} e^{t \L_1} \xii .
\]
Notice that,  if $\norm{\xii}_{H^2 \cap W^{2,1}}\le \e \ll 1$,
then $\xi(t) $ satisfies
\begin{equation*}
\norm{\xi(t)}_{H^2} \le C(n) \e, \quad \norm{\xi(t)}_{W^{2,\infty}}
\le C(n) \e |t|^{-3/2}, \quad \norm{|\xi|^2\xi(t)}_{H^2} \le C(n) \e^3
\bkA{t}^{-3}.
\end{equation*}
Here we have used the boundedness and decay estimates for $e^{t\L_1}\PcL$
in Theorem \ref{th:2-2} (6)--(7). Since $Q_1$ is fixed, it does
not matter that the constant depends on $n$.
The main term of $F_0$ is quadratic in $\xi$.
Hence
\[
\norm{F_0(t)}_{H^2} \le C \e^2 \bkA{t}^{-3}.
\]

As it will become clear, we have the freedom to choose $\xii$ and
$\beta_0=\beta(0)$.
We require that $\xii \in \Hc(\L_1)$ and
\begin{equation}
\norm{\xii}_{H^2 \cap W^{2,1}}\le \e , \qquad |\beta_0|\le \e^2/4,
\end{equation}
with $\e$ sufficiently small.
With given $\xii$ and $\beta_0$, we
will define a contraction mapping $\Tmap$ in the following space
\begin{align*}
{\cal A}=&\big \{ (a, \theta,  \al,\beta, g):
[0, \infty) \to \R \times \R \times \Complex \times \Complex
\times (\Hc(\L_1) \cap H^2),
\\
&\quad |a(t)|, |\al(t)|,|\beta(t)|,
 \le \e^{7/4} (1+t)^{-2} ,
\\
&\quad \norm{g(t)}_{H^2}\le \e^{7/4} (1+t)^{-7/4} ,\quad
 |\theta(t)| \le 2 \e^{7/4} (1+t)^{-1} \ \big \}
\end{align*}
For convenience, we introduce a variable $ b = \dot \theta$.
Our map $\Tmap$ is defined by
\[
\Tmap : (a, \theta, \al, \beta, \eta) \longto (a^\triangle,
\theta^\triangle, \al^\triangle, \beta^\triangle, \eta^\triangle)
\]
where, with  $c_1=  (Q_1,R_1)^{-1}$ and $F=F(aR+h)$ defined in \eqref{F.def},
\begin{align*}
h(t) &= \zeta(t) + \eta(t)
\\
\zeta(t) &=  \RE \bket{ \al(t) \evector +
\beta(t) \sigma_3 \evector} , \qquad
\eta(t)=  U_{\theta(t)}  (e^{ t\L_1} \xii+g(t))
\\
b(t) &=- [a + (c_1 R_1, \, \Re F) ]\cdot
[1+(c_1 R_1, R_1) a +(c_1 R_1,\Re h)]^{-1}
\\
a^\triangle(t) &=  \int_\infty^t
(c_1 Q_1, \, \Re (F + b h)) \,d s
\\
\theta^\triangle(t) &= \int_\infty^t b(s) \,d s
\\
\al^\triangle(t) &= \int _\infty ^t e^{\evalue (t-s)} P_\al
i^{-1} ( F+ b (a R+h))  \, d s
\\
\beta^\triangle(t) &= e^{-\evalue t}\beta_0+\int _0 ^t e^{-\evalue (t-s)}
P_\beta i^{-1} ( F+ b (a R+h)) \, d s
\\
g^\triangle(t) &=\int _\infty ^t e^{ \L_1 (t-s)}
\PcL \Big\{\sin \theta 2 \la Q_1^2 \bar \eta
+ e^{i \theta}  (1-\PcL)i b \eta +
\\
&\qquad \qquad \qquad \qquad +
e^{i \theta} \PcL i^{-1} ( F+ b (a R+\zeta)) \Big\}  \, d s .
\end{align*}

We will use Strichartz estimate for the term
 $\sin \theta 2 \la Q_1^2 \bar \eta$
in the $g$-integral:
\begin{equation} \label{Strichartz}
\norm{\int_\infty^t e^{ \L_1 (t-s)} \PcL f(s,\cdot) \, d s }_{L^2_x}
\le C(n) \bket{\int_\infty^t \norm{f(s,\cdot)}_{L^{r'}_x}^{q'} \, d s
}^{1/{q'}}
\end{equation}
for $\frac 3r + \frac 2q =\frac 32$, $2<q\le \infty$.
Here $\mbox{}'$ means the usual conjugate exponent.
Eq.~\eqref{Strichartz} can be proved by either using wave operator to map
$e^{t\L_1}$ to $e^{-it(-\Delta-E_1)}$, or by using the decay estimate
Theorem \ref{th:2-2} (7) and repeating the usual proof for Strichartz estimate.
We will also use 
\[
\norm{ \phi}_{H^2} \sim \norm{ \L_1 \phi}_{L^2} \qquad \text{for }
\phi \in \Hc(\L_1),
\]
which follows from the spectral gap \eqref{gap2}.
Since $\sin \theta 2 \la Q_1^2 \bar \eta$ is local and bounded by
$C(n) \e^2 (1+t)^{-1}\cdot \e(1+t)^{-3/2}$, by choosing $q$ large we have
\begin{align*}
&\norm{\int_\infty^t e^{ \L_1 (t-s)}
\PcL \sin \theta 2 \la Q_1^2 \bar \eta \, d s }_{H^2}
\\
&\le C(n) \norm{\int_\infty^t e^{ \L_1 (t-s)}
\PcL \L_1\sin \theta 2 \la Q_1^2 \bar \eta \, d s }_{L^2_x}
\\
&\le C(n) \bket{\int_\infty^t  \e^3 (1+s)^{-(5/2)q'} \, d s }^{1/{q'}}
= C(n) \e^3 (1+t)^{-5/2+1/q'}.
\end{align*}
In particular, we get $C\e^3 (1+t)^{-7/4}$ by choosing $q=4$.

Note $|b(t)|\le 2 |a(t)|$.  Since $t-s<0$ in the integrand of $\al$, $\Re
\evalue (t-s)<0$ and the $\al$-integral converges. Similarly $\Re \evalue
(t-s)>0$ in the integrand of $\beta$ and hence the $\beta$-integration
converges. Observe that we have the freedom of choosing $\beta_0$ and
$\xi_\infty$. Since $e^{-\evalue t}\beta_0$ decays exponentially, the main
term of $\beta(t)$ when $t$ large is given by $F_0$, not $e^{-\evalue
t}\beta_0$.  Direct estimates show that
\[
|\al(t)| \le C(n) \e^2 (1+t)^{-3}, \quad
|\beta(t)| \le \e^2 e^{-\gamma t}/4 + C(n) \e^2 (1+t)^{-3},
\]
\[
|a(t)|, |b(t)| \le C(n) \e^2 (1+t)^{-2}, \quad
|\theta(t)|  \le C(n) \e^2 (1+t)^{-1},
\]
\[
\norm{g(t)}_{H^2} \le C(n)  \e^2 (1+t)^{-7/4}.
\]
It is easy to check that the map $\Tmap$ is a contraction
if $\e$ is sufficiently small. Thus we have a fixed point in ${\cal A}$,
which gives a solution to the system \eqref{a.eq}, \eqref{theta.eq},
\eqref{al.eq}, \eqref{beta.eq}, and \eqref{g.eq}. Since it lies in ${\cal
A}$, we also have the desired estimates.  We obtain $\al(0)$, $a(0)$ and
$\theta(0)$ as functions of $\xi_\infty$ and $\beta_0$.

Recall $\psias(t)=Q_1 e^{-i E_1 t+ i \theta(t)} + e^{-i E_1 t} e^{t\L_1}
\xi_\infty$ and we have
\[
\psi(t) = [Q_1 + U_{\theta(t)} e^{t\L_1} \xi_\infty ]\,
e^{-i E_1t+ i \theta(t)} + O(t^{-7/4})  \qquad \text{in }H^2.
\]
Since $\PcL(1-e^{i\theta})=O(\theta(t))=O(t^{-1})$, by
the definition \eqref{solveeta} of $U_\theta$,
\begin{align*}
U_{\theta(t)} e^{t\L_1} \xi_\infty
&=[1+\PcL(1-e^{i\theta})]e^{t\L_1} \xi_\infty  + O(t^{-2})
\\
&=
(2-e^{i\theta})e^{t\L_1} \xi_\infty +
(1-\PcL)(1-e^{i\theta})e^{t\L_1} \xi_\infty + O(t^{-2})
\end{align*}
in $H^2$. Since $(1-\PcL)$ is a local operator, 
$(1-\PcL)(1-e^{i\theta})e^{t\L_1} \xi_\infty =O(t^{-1} \cdot t^{-3/2})$.
Also, $e^{i\theta}(2-e^{i\theta}) = 1+ O(\theta^2) = 1 + O(t^{-2})$.
Hence we have $\psi(t)-\psias(t) = O(t^{-7/4})$ in $H^2$.
We have proven Theorem \ref{th:1-2} under assumption (R).

\bigskip

We now sketch the proof for the non-resonant case.
The only difference is that we define $\zeta(t)$ as
$\RE \al(t) \evector $ and write $\psi(t)$ in the form
\[
\psi(t) =  \bket{ Q_1 + a(t)R_1+ \RE \bke{ \al(t) \evector }
+ U_{\theta(t)} (e^{ t\L_1} \xii+g(t) ) } \, e^{-i E_1 t+ i \theta(t)}.
\]
The function $\al(t)$ still satisfies \eqref{al.eq} 
but with a purely imaginary eigenvalue $\evalue$. The previous proof
will go through if we remove all terms related to $\beta$.

\section{Appendix: Vanishing solutions}

In this appendix we prove Proposition \ref{th:1-1}.
Recall $H_0=- \Delta +V$. The propagator $e^{-iH_0t}$ is bounded
in $H^s$, $s\ge 0$, and satisfies the decay estimate of the form
\eqref{decayest}:
\begin{equation} \label{decayest0}
\norm{ e^{-itH_0} \Pc^{H_0} \phi}_{L^\infty} \le C |t|^{- 3/2}
\norm{\phi}_{L^1}
\end{equation}
under our assumption A1. See \cite{JK,JSS,Y}.

For any $\xii \in \Hc(H_0)$ with small $H^2\cap W^{2,1}$ norm, we
want to construct a solution $\psi(t)$ of \eqref{Sch} with the
form
\begin{equation} \label{2-1}
\psi(t) = e^{-iH_0t} \xii + g(t), \qquad g(t)=\text{error}.
\end{equation}
Let $\xi(t) =  e^{-iH_0t} \xii$. Suppose $\norm{\xii}_{H^2\cap
W^{2,1}} = \e$, $0< \e\le \e_0$, we have by \eqref{decayest0},
\[
 \norm{\xi(t)}_{H^2} \le C_1 \e,
   \norm{\xi(t)}_{W^{2, \infty}} \le C_1 \e |t|^{-3/2}, \quad
   \norm{|\xi|^2\xi(t)}_{H^2} \le  C_1 \e^3 (1+t)^{-3}
\]
for some constant $C_1$.

The error term $g(t)$ satisfies
\[
\pd_t g = -i H_0 g + F
\]
with $g(t) \to 0$ as $t \to \infty$ in certain sense, and
\begin{equation} \label{2-2}
F(t)= -i \la |\psi|^2 \psi
= -i \length{ \xi(t) + g(t)}^2 ( \xi(t) + g(t)) ~, \quad
\xi(t)=e^{-iH_0t} \xii.
\end{equation}
We define a solution by \eqref{2-2} and
\begin{equation}
g(t) = \int_\infty^t  e^{-iH_0 (t-s)}  F(s)  \, d s ~.
\end{equation}
Note that our $g(t)$ belongs to $L^2$ and is not restricted to the
continuous spectrum component of $H_0$. Also note that the main
term in $F$ is $|\xi|^2 \xi(t)$, which is of order $t^{-3}$ in
$H^2$. Hence $g(t) \sim t^{-2}$.

We define a contraction mapping in the following class
\[
\mathcal{A} = \bket{ g(t):[0,\infty) \to H^2(\R^3), \,
 \norm{h(t)}_{H^2} \le C_1 \e^3(1+t)^{-2} } ~ .
\]
This class is not empty since it contains the zero function.
We also define the norm
\[
\norm{g}_{\mathcal A} := \sup _{t>0} (1+t)^2 \norm{g(t)}_{H^2}.
\]
For $g(t) \in \mathcal{A}$ we define
\[
\Tmap: g(t) \to
g^\triangle(t) =-i\la \int_\infty^t  e^{-iH_0 (t-s)}
 \bke{|\xi + g|^2 (\xi + g)} (s)  \, d s ~.
\]
It is easy to check that
\begin{align*}
\norm{g^\triangle(t)}_{H^2} &\le \int_t^\infty \norm{F(t)}_{H^2} \,
d s
\\
&\le \int_t^\infty  C_1 \e^3 \bkA{s}^{-3} + C \e^5
{\bkA{s}^{-7/2}}  \, d s \le C_1 \e^3\bkA{t}^{-2} ~,
\end{align*}
if $\e_0$ is sufficiently small. This shows that the map $\Tmap$
maps $\mathcal{A}$ into itself. Similarly one can show
$\norm{\Tmap g_1 - \Tmap g_2}_{\mathcal A} \le \norm{ g_1 -
g_2}_{\mathcal A}$, if $g_1, g_2 \in \mathcal{A}$. Therefore our
map $\Tmap$ is a contraction mapping and we have a fixed point.
Hence we have a solution $\psi(t)$ of the form \eqref{2-1} with
$e^{-itH_0}\xii$ as the main profile.

{\bf Remark.} The above existence result holds no matter how many
bound states $H_0$ has.  The situation is different if we
linearize around a nonlinear excited state. In that case, the
propagator $e^{t\L_1}$, ($\L_1$ is the linearized operator), may
not be bounded in whole $L^2$.

\subsection*{Acknowledgments}

The authors would like to thank L. Erd\"os for his very helpful comments
and discussions. Part of this work was done when both authors visited the
Academia Sinica and the Center for Theoretical Sciences in Taiwan. The
hospitalities of these institutions are gratefully acknowledged.
Tsai was partially supported by NSF grant DMS-9729992.
Yau was partially supported by NSF grant DMS-0072098.

\bigskip

\bigskip

\noindent{Tai-Peng Tsai},  ttsai@ias.edu \\
Institute for Advanced Study, Princeton, NJ 08540

\bigskip

\noindent{Horng-Tzer Yau}, yau@cims.nyu.edu \\
Courant Institute, New York University, New York, NY 10012

\end{document}